\documentclass[%
  reprint,
superscriptaddress,
nofootinbib,
 amsmath,amssymb,
 aps,
 prc,
 showkeys,
]{revtex4-1}

\usepackage{graphicx}
\usepackage{dcolumn}
\usepackage{bm}
\usepackage{subfigure}
\usepackage{float} 
\usepackage{hyperref}
\usepackage[mathlines]{lineno}
\usepackage{lipsum, babel}

\newcommand{\be}{\begin{equation}}
\newcommand{\ee}{\end{equation}}
\newcommand{\bea}{\begin{eqnarray}}
\newcommand{\eea}{\end{eqnarray}}

\begin{document}

\preprint{BeAGLE-BNL}

\title{Investigation  of the background in coherent $J/\psi$ production at the EIC}
\author{Wan~Chang}
\email{changwan@mails.ccnu.edu.cn}
\affiliation{%
 Key Laboratory of Quark and Lepton Physics (MOE) and Institute of Particle Physics, Central China Normal University, Wuhan 430079,China
}%
\affiliation{%
 Department of Physics, Brookhaven National Laboratory, Upton, NY 11973, U.S.A.
}%
\author{Elke-Caroline~Aschenauer }
\email{elke@bnl.gov}
\affiliation{%
 Department of Physics, Brookhaven National Laboratory, Upton, NY 11973, U.S.A.
}%
\author{Mark~D.~Baker}
 \email{mdbaker@mbdpads.com}
\affiliation{%
 Mark D. Baker Physics and Detector Simulations LLC, Miller Place, NY 11764, U.S.A.
}%
\author{Alexander~Jentsch}
\email{ajentsch@bnl.gov}
\affiliation{%
 Department of Physics, Brookhaven National Laboratory, Upton, NY 11973, U.S.A.
}%
\author{Jeong-Hun~Lee}
\affiliation{%
 Department of Physics, Brookhaven National Laboratory, Upton, NY 11973, U.S.A.
}%
\author{Zhoudunming~Tu}
\email{zhoudunming@bnl.gov}
\affiliation{%
 Department of Physics, Brookhaven National Laboratory, Upton, NY 11973, U.S.A.
}%
\affiliation{Center for Frontiers in Nuclear Science, Stony Brook, NY 11794, U.S.A.}
\author{Zhongbao~Yin}
\affiliation{%
 Key Laboratory of Quark and Lepton Physics (MOE) and Institute of Particle Physics, Central China Normal University, Wuhan 430079,China
}%
\author{Liang~Zheng}
\affiliation{
 School of Mathematics and Physics, China University of Geosciences, Wuhan 430074, China
}%

\date{\today}

\begin{abstract}
Understanding various fundamental properties of nucleons and nuclei are among the most important scientific goals at the upcoming Electron-Ion Collider (EIC). With the unprecedented opportunity provided by the next-generation machine, the EIC might provide definitive answers to many standing puzzles and open questions in modern nuclear physics. Here we investigate one of the golden measurements proposed at the EIC, which is to obtain the spatial gluon density distribution within a lead ($Pb$) nucleus. The proposed experimental process is the exclusive $J/\psi$ vector-meson production off the $Pb$ nucleus - $e+Pb\rightarrow e'+J/\psi+Pb'$. The Fourier transformation of the momentum transfer $|t|$ distribution of the coherent diffraction is the transverse gluon spatial distribution. In order to measure it, the experiment has to overcome an overwhelmingly large background arising from the incoherent diffractive production, where the nucleus $Pb'$ mostly breaks up into fragments of particles in the far-forward direction close to the hadron-going beam rapidity. In this paper, we systematically study the rejection of incoherent $J/\psi$ production by vetoing products from these nuclear breakups - protons, neutrons, and photons, which is based on the BeAGLE event generator and the most up-to-date EIC Far-forward Interaction Region design. The achieved vetoing efficiency, the ratio between the number of vetoed events and total incoherent events, ranges from about 80\% - 99\% depending on $|t|$, which can resolve at least the first minimum of the coherent diffractive distribution based on the Sar\textit{t}re model. Experimental and accelerator machine challenges as well as potential improvements are discussed. 

\end{abstract}

\maketitle


\section{\label{sec:intro}Introduction}
The future US-based Electron Ion Collider (EIC)~\cite{ref:EICCDR} will be capable of colliding electrons with a range of ions - from protons to uranium - over a broad range of energies, and at very high luminosity ($10^{33-34} \rm{cm}^{-2} \rm{sec}^{-1}$). Additionally, the EIC will enable collisions of polarized electrons and light ions ($p$, $^{3}He$), being the only collider in the world with this capability. The EIC will enable study of Quantum Chromodynamics (QCD) and the imaging of the quarks and gluons, and their interactions in QCD, at a previously unattainable level of precision ~\cite{Accardi:2012qut}. The EIC will open up the unique opportunity to go far beyond the present one-dimensional picture of nuclei and nucleons, where the composite nucleon appears as many fast-moving (anti-)quarks and gluons whose transverse momenta or spatial extent are not resolved. Specifically, correlating the information of the longitudinal momentum of quarks and gluons with their transverse momentum and spatial distribution inside the nucleon will enable nuclear “femtography”. Such femtographic images will provide, for the first time, insight into the QCD dynamics inside hadrons, such as the interplay between sea quarks and gluons. Investigating gluons in nuclei instead of protons has multiple advantages. Namely, that nuclei act as an effective ``amplifiers" of phenomena related to high gluon densities, which enhance the impact of nonlinear gluon interactions which possibly lead to gluon saturation~\cite{Munier:2001nr}, also known as the Colour Glass Condensate~\cite{Gelis:2010nm, Jalilian-Marian:2014ica, Jalilian-Marian:2005ccm, Weigert:2005us, Iancu:2003xm}. The EIC has the potential to map the transition from a linear to a nonlinear regime in QCD and characterize the relevant parameters governing this transition.

One of the golden measurements proposed at the EIC is the detection of coherent and incoherent vector-meson (VM) production from heavy nuclei~\cite{Accardi:2012qut}. This measurement has three important physics implications. Coherent production is: i) a direct measurement of the parton spatial distribution inside of a nucleus; ii) sensitive to non-linear dynamics in QCD~\cite{Kowalski:2003hm, Accardi:2012qut, Toll:2012mb, Sambasivam:2019gdd} when comparing the production of different VMs in different kinematic regions. iii) according to the Good-Walker picture~\cite{PhysRev.120.1857}, the incoherent cross-section is a direct measure of the lumpiness of the gluon distribution in the ion.

Similar to single-slit diffraction in optical experiments, the coherent diffractive production of vector mesons in high-energy experiments is directly sensitive to the size of the target. 
The most promising channel to map the spatial gluon distribution in nuclei is to measure coherent $J/\psi$ production off a heavy nucleus, such as lead ($Pb$),  where the scattered $Pb$ nucleus is required to stay intact as described in the reaction process, $e+Pb\rightarrow e'+J/\psi+Pb'$.
The gluon density distribution in transverse impact-parameter space is related by a Fourier transformation with the distribution of the momentum transfer $|t|$~\cite{Toll:2012mb} as follows, 
\be
F(b)=\frac{1}{2\pi}\int^{\infty}_{0}d\Delta \cdot \Delta J_{0}(\Delta b)\sqrt{\frac{d\sigma_{\rm coherent}}{d|t|}(\Delta)} \biggr\rvert_{\rm{mod}}.
\ee
\noindent Here $F(b)$ is the gluon density distribution as a function of impact parameter $b$, $\Delta=\sqrt{-t}$, $J_{0}$ is the Bessel function, and $d\sigma_{\rm{coherent}}/d|t|$ is the coherent differential cross section. It is critical for the proposed measurement to unambiguously identify the coherent process and measure its differential cross section as a function of $|t|$. In addition, for the same gluon density distribution in a saturated regime, using different VM probes may result in different coherent cross sections, leading to different measured gluon density distributions~\cite{Toll:2012mb}. However, given the mass of the $J/\psi$ particle, the predicted sensitivity to saturation effects in coherent $J/\psi$ production is found to be negligible~\cite{Sambasivam:2019gdd}, which makes it an ideal baseline to identify gluon saturation effects. In order to probe saturation dynamics, e.g. its impact on gluon distributions, comparison of $\phi$ meson with respect to the $J/\psi$ in coherent production will be essential. 

\begin{figure}[thb]
\centering
\includegraphics[width=3.3in]{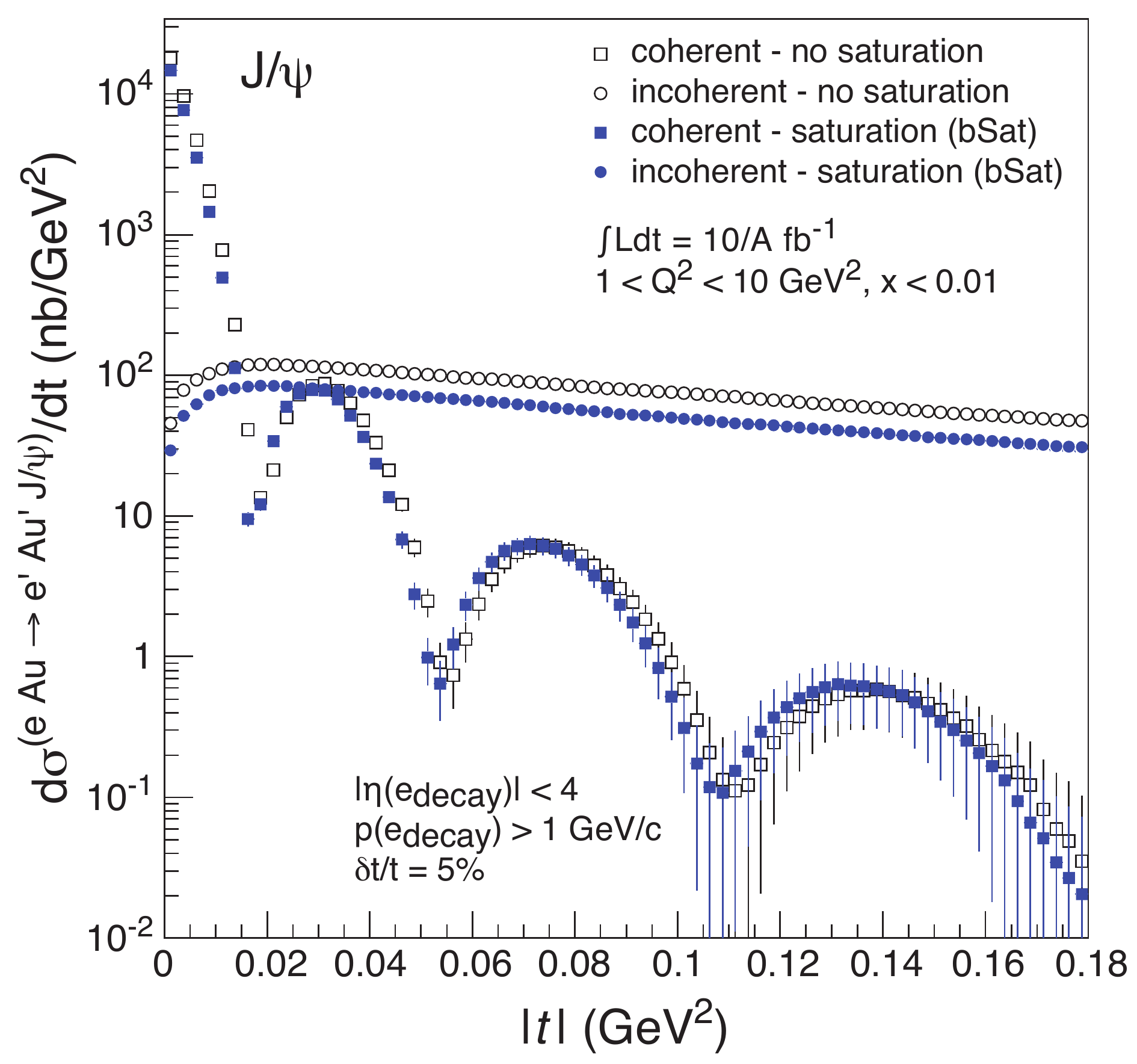}
  \caption{ \label{fig:figure_diffractive} Differential cross-sections of $J/\psi$ production in $eAu$ collisions at the EIC based on the Sar\textit{t}re model~\cite{Toll:2012mb}. Both coherent and incoherent production with and without saturation effect are shown.}
\end{figure}

However, the competing process of exclusive incoherent vector meson production, $e+Pb\rightarrow e^{'}+J/\psi+X$, occurs when the primary interaction takes place at the nucleon level instead of the nucleus. The nucleus could be then broken up by the virtual photon (or the color dipole~\cite{Kowalski:2003hm}) into nuclear remnants and nucleons, where individual nucleons stay intact with very small scattering angles. Since the target size between a $Pb$ nucleus ($\sim 8~\rm{fm}$) and a nucleon ($\sim 0.8~\rm{fm}$) is different by one order of magnitude, the resulting distribution of the momentum transfer $|t|$ is expected to be drastically different. The slope of the $|t|$ distribution at low $|t|$ is expected to be inversely proportional to the gluon transverse size~\cite{Toll:2012mb,Klein:2016yzr}, where for $|t|> 0.015~\rm GeV^{2}$ the incoherent contribution starts to dominate, as shown in Fig.~\ref{fig:figure_diffractive} for the Sar\textit{t}re model simulation~\cite{Toll:2012mb,Toll:2013gda}. Note that the Sar\textit{t}re model uses the gold ($Au$) nucleus instead of $Pb$, while the difference between the two nuclei is expected to be small compared to other uncertainties\footnote{$ePb$ collisions are not currently available in the released version of the Sar\textit{t}re model simulation.}. From a recent quantitative study in the EIC Yellow Report~\cite{AbdulKhalek:2021gbh}, resolving the three diffractive minima of the coherent $|t|$-distribution is critical to achieve the goal of this measurement with reasonable precision on the gluon density distribution. In order to observe the three minima from low to high $|t|$ at the EIC, the required vetoing efficiency, the ratio between number of vetoed events and total incoherent events, is roughly 90\%, 99\%, and $>99.8\%$, respectively. The three diffractive minima are also shown in Fig.~\ref{fig:figure_diffractive}, where a 5\% resolution effect was included in generating the $|t|$ distributions. 

Similarly, at the Relativistic Heavy-Ion Collider (RHIC) and the Large Hadron Collider (LHC), the effort of measuring the gluon density distributions of heavy nuclei has already begun using photoproduction of $\rho^{0}$ and $J/\psi$ mesons in ultra-peripheral collisions (UPCs)~\cite{Khachatryan:2016qhq,Abelev:2012ba,Adam:2015gsa,Adamczyk:2017vfu,Acharya:2020sbc,ALICE:2021jnv,ALICE:2021tyx,ALICE:2021gpt}. These measurements are very similar to the proposed measurement at the EIC, except that the photon virtuality is close to zero in UPCs and one has less control on the event-by-event kinematics. An experimental hint of coherent production has been observed in photoproduction of $\rho^{0}$ mesons at RHIC using the STAR detector in $AuAu$ UPC events~\cite{Adamczyk:2017vfu}. But the data of the coherent distribution are strongly smeared by the photon transverse momentum and incoherent contributions. These results therefore can not be used to extract a precise gluon density distribution~\cite{Adamczyk:2017vfu}. Rejecting incoherent contributions in UPC data with an event-by-event experimental method is extremely challenging, if not impossible, since the Zero-Degree Calorimeter (ZDC) is the only available far-forward detector at STAR during heavy-ion collisions. Therefore, the UPC data at RHIC and at the LHC, as of now, cannot achieve the goal of measuring the gluon density distribution in a heavy nucleus, while the planned EIC-experiments with their unique detector capabilities along the beam-line might have the best opportunity in fulfilling this experimental quest in the future.

In this paper, we characterize the dominant background contribution of the coherent $J/\psi$ vector meson production in $ePb$ collisions at the top EIC energy $\sqrt{s}=89~\rm{GeV}$. Specifically, we use the BeAGLE event generator to simulate incoherent diffractive $J/\psi$ events, where each event consists of the produced $J/\psi$ at mid-rapidity, and all the particles from the nuclear breakup at forward-rapidities. The final-state particles produced by the nuclear breakup can be any combination of protons, neutrons, photons, and nuclear remnants. These BeAGLE Monte Carlo (MC) events are processed in a GEANT-based simulation of the EIC interaction region (IR) and its far-forward (FF) detectors in order to investigate their acceptances and the impact of beam-related effects. Since the BeAGLE generator does not predict the coherent $J/\psi$ vector meson production~\cite{Beagle}, we adopt the ratio of the incoherent to coherent cross section from the Sar\textit{t}re model to define the position and relative magnitude of the three diffractive minima. An event is considered vetoed if there is at least one particle detected in any of the FF detectors. The goal of this study is to see how many events can be vetoed given the current IR and FF detectors using the most up-to-date full detector simulations. The results will provide valuable insight to the detector proposal and future IR improvements at the EIC.   

This paper is organized as follows. In Sec.~\ref{sec:generator}, the event generator BeAGLE will be briefly introduced. In Sec.~\ref{sec:detectors}, the forward detectors along the outgoing hadron beam will be discussed. In Sec.~\ref{sec:result}, the final results will be shown, followed by a discussion of the remaining issues and challenges faced by this measurement in Sec.~\ref{sec:discussions}. Finally, a summary will be given in Sec.~\ref{sec:summary}.

\section{\label{sec:generator} BeAGLE}
BeAGLE is a general-purpose electron-nucleus event generator for high energy $eA$ collisions. It has been extensively used to understand $eA$ physics and the EIC detector/interaction region design~\cite{AbdulKhalek:2021gbh}. The core of the BeAGLE model is based on the PYTHIA-6 event generator~\cite{Sjostrand:2006za} for simulating the parton level interactions in electron-nucleon collisions. The nuclear geometry is modeled within a Glauber-type formalism. Final-state interactions between produced particles and spectator nucleons are provided by the program of DPMJET~\cite{Roesler:2000he}. Finally, the FLUKA model~\cite{Bohlen:2014buj,Ferrari:2005zk} is implemented to describe the breakup of the excited nucleus. Below, only a few important features of the BeAGLE model that are most relevant to this paper are described.  For details, see Refs.~\cite{Beagle,Tu:2020ymk}. 

BeAGLE uses a Woods-Saxon distribution  for nucleons in heavy nuclei. A Glauber-type multiple scattering formalism is applied to the scattering on the nuclear target, although only one hard $\gamma^{\ast}+N$ interaction per event is allowed.  Final state interactions are modeled through the DPMJET~\cite{Roesler:2000he} intra-nuclear cascade (INC) process~\cite{PhysRev.131.1801} , which describes the secondary interactions among the spectator nucleons and the products from the primary electron nucleon scattering. This is implemented using a formation time $\tau$: the average time needed for creating a secondary particle that might interact with other nucleons. The average hadron formation time $\tau$ is defined as follows~\cite{Ferrari:1995cq, Zheng:2014cha}, 
\begin{equation}  
\tau =\tau_{0}\frac{E}{m}\frac{m^{2}}{m^{2}+p_{\perp}^{2}},
\label{equation:formationtime} 
\end{equation}
\noindent where $E$, $m$ and $p_{\perp }$ are the energy, mass and transverse momentum of the secondary particle, respectively. For generating a secondary particle, a particular formation time $T$ is randomly sampled from an exponential distribution, $e^{-T/\tau}$, where the average formation time is set by the $\tau$ parameter in Eq.~(\ref{equation:formationtime}). Hadrons with higher energy or smaller transverse mass are more likely to have a longer formation time and less likely to be formed inside the nucleus.  Note that the parameter $\tau_{0}$ is a free parameter, which has been tuned by comparing to experimental data. The observable that was used, which is sensitive to the $\tau_{0}$ parameter, is the average number of neutrons, $\langle  N_{n} \rangle$, produced from the evaporation process. We tuned to the multiplicity data of neutron emission in $\mu Pb$ collisions from the E665 experiment at Fermilab~\cite{E665:1995utr}. 

 BeAGLE does not simulate coherent diffraction from the entire nucleus, while the E665 data does have coherent diffractive events, which do not produce neutrons in the final state. In order to properly tune the $\tau_{0}$ parameter, a weight is needed for the BeAGLE model to take into account the coherent cross section in the data. The comparison of BeAGLE to the E665 data requires the introduction of the fraction of coherent events over the total cross section $f=N_{\rm{coherent}}/N_{\rm{total}}$ as follows,
\be
N_{n}({\rm E665}) = 0* f + N_{n}({\rm BeAGLE})*(1-f).
\label{equation:E665}
\ee

\noindent By using this relation, we find the $\tau_{0}=6$~fm for $f=0.42$, 10~fm for $f=0.24$, and 14~fm for $f=0.08$. The default BeAGLE $\tau_{0}$ is set for 10~fm, while the other two are served as model uncertainty in estimating the rejection power in the incoherent events. See details in Appendix~\ref{app:tau0}. 

In addition, two aspects of nuclear shadowing~\cite{Frankfurt:1988nt} have been implemented  in BeAGLE. First, the cross-section for all of the different hard interactions is affected by a modification in the parton distribution for the bound nucleons. Second, multi-nucleon scattering according to a Glauber model is available in BeAGLE, with three possible settings.  For the first setting, i.e., \textit{genShd} = 1, it is assumed that one and only one nucleon participates in the hard interaction with the virtual photon. For \textit{genShd} = 3, the photon interacts with multiple nucleons and only one of the struck nucleons is selected randomly to undergo the inelastic interaction. The rest of photon-nucleon interactions are treated as elastic. For \textit{genShd} = 2, the process is the same as \textit{genShd} = 3 except the order is fixed in a way that the first interaction is always inelastic and the rest is elastic. For tuning the $\tau_{0}$ parameter, the \textit{genShd} is set to 2. Using different settings in the nuclear shadowing model has a negligible effect on the E665 evaporation neutron results. 

Finally, for incoherent $J/\psi$ production from nuclei, e.g., $Pb$, the primary interaction is based on the electroproduction of $J/\psi$ as modeled in PYTHIA-6. The active nucleon in the BeAGLE event generator can be either a proton or neutron. For the case of the neutron, the BeAGLE model assumes isospin symmetry for the parton distributions. BeAGLE uses nuclear parton distribution functions (nPDFs) for the basic electron-nucleon hard cross-section. For these results, the EPS09 parameterization~\cite{Eskola:2009uj} was used.  After the hard interaction, the leading nucleon can stay intact (elastic on nucleon level) or dissociate (inelastic on nucleon level) given by two separate processes modeled in PYTHIA-6, namely subprocess 91 and 93, respectively. 

In the measurement of coherent $J/\psi$ production at the EIC, these two processes of incoherent production are the main contributions of the physics background. The resulting final-state products in these processes are therefore the main focus of this paper, which will be mostly produced in the very forward pseudo-rapidity region at the EIC. For a similar process in light nuclei, e.g.~deuterons, see Ref.~\cite{Tu:2020ymk} for details.

The BeAGLE simulation used in this paper is based on a sample of $e+Pb\rightarrow e^{'}+J/\psi+X$ with 18 GeV electrons scattering off 110 GeV per nucleon $Pb$ nuclei. 1.3 million events of incoherent $J/\psi$ production have been simulated. 
\section{\label{sec:detectors} Far-Forward Detectors}
The current EIC IR and far-forward region design are based on the EIC Conceptual Design Report (CDR)~\cite{ref:EICCDR}. The IR used in this study is located at the 6 o'clock position (IP6) of the current Relativisitic Heavy Ion Collder (RHIC) complex at BNL, which is the current location of the STAR detector. The FF detectors considered in this study are advanced concepts for measuring forward-going particles that are outside the main detector acceptance ($\theta > $ 35 mrad), and are based on the EIC reference detector detailed in the EIC Yellow Report~\cite{AbdulKhalek:2021gbh}. Some general considerations used to establish baseline particle acceptances and detector resolutions for the present study via full simulations in EICRoot~\cite{ref:EICROOT} and Geant~\cite{Brun:1987ma, GEANT4} are presented here.

\begin{table*}[thb]
\fontsize{11}{13}\selectfont
\begin{center}
\begin{tabular}{|c|c|c|c|c|}
\hline
  \textbf{Detectors} & ($x$,$z$) Positions [m] & Dimensions  & $\theta$ [mrad] & Notes \\
\hline
 B0 tracker & ($x=0.19$, $5.4~<~z<~~6.4$) & (26 cm, 27 cm, n/a ) & $5.5<\theta<20$ & ~13 mrad at $\phi=\pi$ \\
 Off-momentum & (0.8, 22.5),(0.85,24.5)  & (30 cm, 30 cm, n/a) & $0<\theta<5.0$ & 40\% to 60\% rigidity \\
 Roman Pots & (0.85, 26.0), (0.94, 28.0) & (25 cm, 10 cm, n/a) & $0<\theta<5.0$ & 10$\sigma$ cut \\
 ZDC &  (0.96, 37.5) & (60 cm, 60 cm, 2~m) & $0<\theta<5.5$ & $\sim$ 4.0 mrad at $\phi = \pi $ \\
\hline
 \end{tabular}
 \end{center}
  \caption{\label{tab:table1} Summary of the physical location, maximum polar angular acceptance for the four far-forward detectors~\cite{AbdulKhalek:2021gbh}.}
 \end{table*}

\begin{figure}[tbh]
\includegraphics[width=\linewidth]{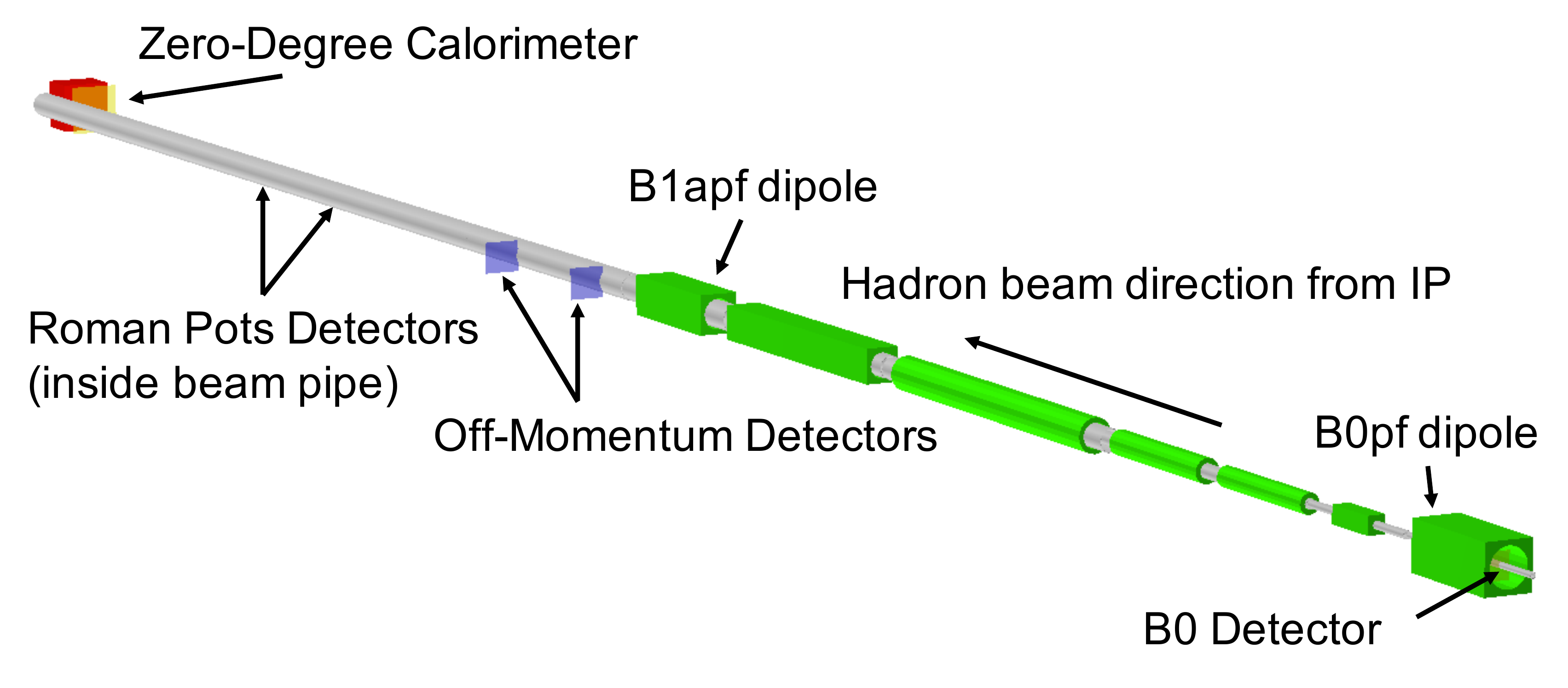}
  \caption{\label{fig:figure_IR}The layout of the EIC far-forward IR region including the seven dipole and quadrupole magnets for the outgoing hadron beam direction and the four detector subsystems for proton and neutron tagging. Image generated using EicRoot~\cite{ref:EICROOT} and GEANT4~\cite{GEANT4}}
\end{figure}

In Fig.~\ref{fig:figure_IR}, the layout of the EIC far-forward region at IP6 is shown. The dipole magnets are represented by the green rectangular boxes; the quadrupole magnets, which are used to focus the hadron beams, are represented by the green cylindrical boxes. The gray cylindrical tube represents a simplified preliminary estimate for the beam pipe. In order to tag the maximum number of final-state particles in the FF region, four detector subsystems are employed, each with a different area of geometric acceptance coverage. The FF region, in general, covers a polar angle acceptance of 0 to 20 mrad, and provides coverage for particles of different rigidity (momentum divided by charge) compared to the beam (so-called ``off-momentum" particles). The position, geometric dimensions, and angular acceptances of these far-forward detectors are summarized in Table.~\ref{tab:table1}.

Starting from the interaction point and moving downstream, the first detector sub-system is the the B0 spectrometer. The B0 system comprises four silicon tracking layers spaced evenly along the majority of the bore length and allows for tagging and reconstruction of charged particles in a $\sim$1.3 T dipole field, enabling reconstruction of particle momentum. Additionally, a pre-shower detector, consisting of a 11.6 mm (2 radiation length) thick lead layer and a 0.3 mm silicon layer, was included in the B0 magnet bore and after the silicon tracking planes. The entire B0 spectrometer subsystem is designed to tag particles with scattering angles between $\sim$~6 and 20 mrad, with the lower bound determined by the size of the hadron beam pipe in the magnet bore ($r = 3.2$~cm), and the upper bound determined by the aperture (bore size) of the magnet.

The next detector subsystem is the so-called off-momentum detector (OMD). In the present study, the layout of the OMD was identical to that used in the EIC Yellow Report.  The OMD comprises two silicon planes, spaced two meters apart, and placed just after the B1apf dipole magnet outside the beam pipe vacuum. This subsystem is responsible for capturing protons (or other charged particles) from nuclear breakup, which have a lower magnetic rigidity compared to the heavy-ion beam. The lower magnetic rigidity causes these charged particles to experience a larger bending angle in the dipole magnets, causing them to be steered out of the beam pipe. Some further optimizations of the off-momentum detectors are currently under consideration, but were not available at the time of this study. The OMD subsystem covers a polar angle acceptance of 0 to 5 mrad, and accepts particles roughly between 30\% and 60\% of the beam rigidity.

 The third detector subsystem is the Roman Pots (RP) silicon detectors. The RP subsystem consists of two stations of silicon planes, spaced two meters apart, which are injected directly into the beam vacuum a few millimeters from the hadron beam. In other implementations of the RP concept, a metal vessel (the ``pot") with a thin window is used to house the detector packages such that the vessel sits in the machine vacuum, while the silicon detectors are outside the vacuum and protected by the pot. In order to maximize the acceptance at lower transverse momentum, the simulations used in the present study assume a ``potless" design, with the silicon detectors placed directly in the machine vacuum. The RP system nominally accepts particles between 0 and 5 mrad, and a with a rigidity between 60\% and 100\%. However, the detector cannot be placed arbitrarily close to the beam. The safe distance is generally defined as a ``rule-of-thumb" 10$\sigma$ distance, where $\sigma$ is the transverse beam size and is calculated based on the emittance ($\epsilon$), beta-functions ($\beta$), momentum dispersion ($D$), and momentum spread ($\Delta p/p$) of the beam, as shown in Eq.~\ref{eq:beamsize},
 
 \begin{equation}
     \sigma_{x,y} = \sqrt{\epsilon_{x,y}\beta(z)_{x,y} + D_{x,y}\frac{\Delta p}{p}}.
     \label{eq:beamsize}
 \end{equation}
 
 For the beam conditions detailed in the EIC CDR~\cite{ref:EICCDR} and used in the present simulations, 1$\sigma$ is a few millimeters in x, and less than a millimeter in y. The transverse beam size at the RP detectors can be altered using different beam optics configurations, which trade-off acceptance at the detector and overall luminosity.
 
The final detector subsystem is the Zero-Degree Calorimeter (ZDC), which sits just before the B2pf dipole magnet at the end of the drift region. Note the B2pf dipole magnet is not shown in Fig.~\ref{fig:figure_IR}. The full detector will include both electromagnetic and hadronic calorimetry with high resolution ($\frac{\Delta E}{E} = \frac{50\%}{\sqrt{E}} \oplus 5\%$ and $\frac{\Delta\theta}{\theta} = \frac{3\rm{mrad}}{\sqrt{E}}$, where $E$ represents energy deposition in unit of GeV) facilitated by high granularity and large length (2 meters) for shower development. The ZDC used for this study did not have the full implementation included, and instead uses a plane for detecting the generic neutral particle acceptance, with the resolutions applied as an after-burner to the generator-level particle kinematics. The ZDC can accept neutrons and photons with polar angles between 0 and 4.5 mrad.

The beam pipe used in these simulations is based on a preliminary design with basic assumptions and minimal optimization. The final, optimized vacuum engineering design for the first EIC interaction region (IP6) is still underway at the time of this publication. Multiple beam pipe materials (beryllium, aluminum, and stainless steel) were considered for their impact on veoting efficiency, which will provide crucial input to the engineers designing the final vacuum system.

The coordinate system is defined with the z-axis in the hadron-going beam direction, the x-axis determines the position along the floor transverse to the beam (with positive x following the direction of the dipole bending), and the y-axis is the elevation. All detector and beam-lattice components are at the same elevation (i.e. same y-coordinate), with the details being available in the EIC CDR~\cite{ref:EICCDR}.

\section{\label{sec:result} Result}
In this section, we investigate the background contribution to exclusive coherent $J/\psi$ production from the incoherent process, i.e., $e+Pb\rightarrow e^{'}+J/\psi+X$. In fact, this process is the dominant background contribution for most of the range of the momentum transfer $|t|$. In order to suppress the incoherent process, the final-state particles would have to be detected event-by-event using forward particle detectors. In the BeAGLE model, the incoherent $J/\psi$ is produced together with one or more ions and, as shown in Table~\ref{tab:table_1}, either protons, neutrons, photons, or any combination of them depending on the excitation energy. Most events have more than one such particle produced, while around 10\% of events only have either neutrons or photons. Only one particle is needed to be detected for a successful veto, so naturally events with fewer particles are generally more difficult to veto.

\begin{table}[h]
\fontsize{11}{13}\selectfont
\centering
\begin{tabular}{c|c}
\hline
produced particle & rate \\\hline
only neutron & 7.66\% \\\hline
only proton & 0\% \\\hline
only photon & 3.25\% \\\hline
neutron and proton & 3.19 \% \\\hline
neutron and photon & 44.24 \% \\\hline
proton and photon & 2.27 \% \\\hline
neutron, proton and photon & 39.39 \% \\\hline
\end{tabular}
\caption{Summary of particles produced in incoherent $J/\psi$ production in BeAGLE.}
\label{tab:table_1}
\end{table}

\begin{figure*}[thb]
\centering
\subfigure[]{
\label{fig:figure_4_a}
\includegraphics[width=0.3\textwidth]{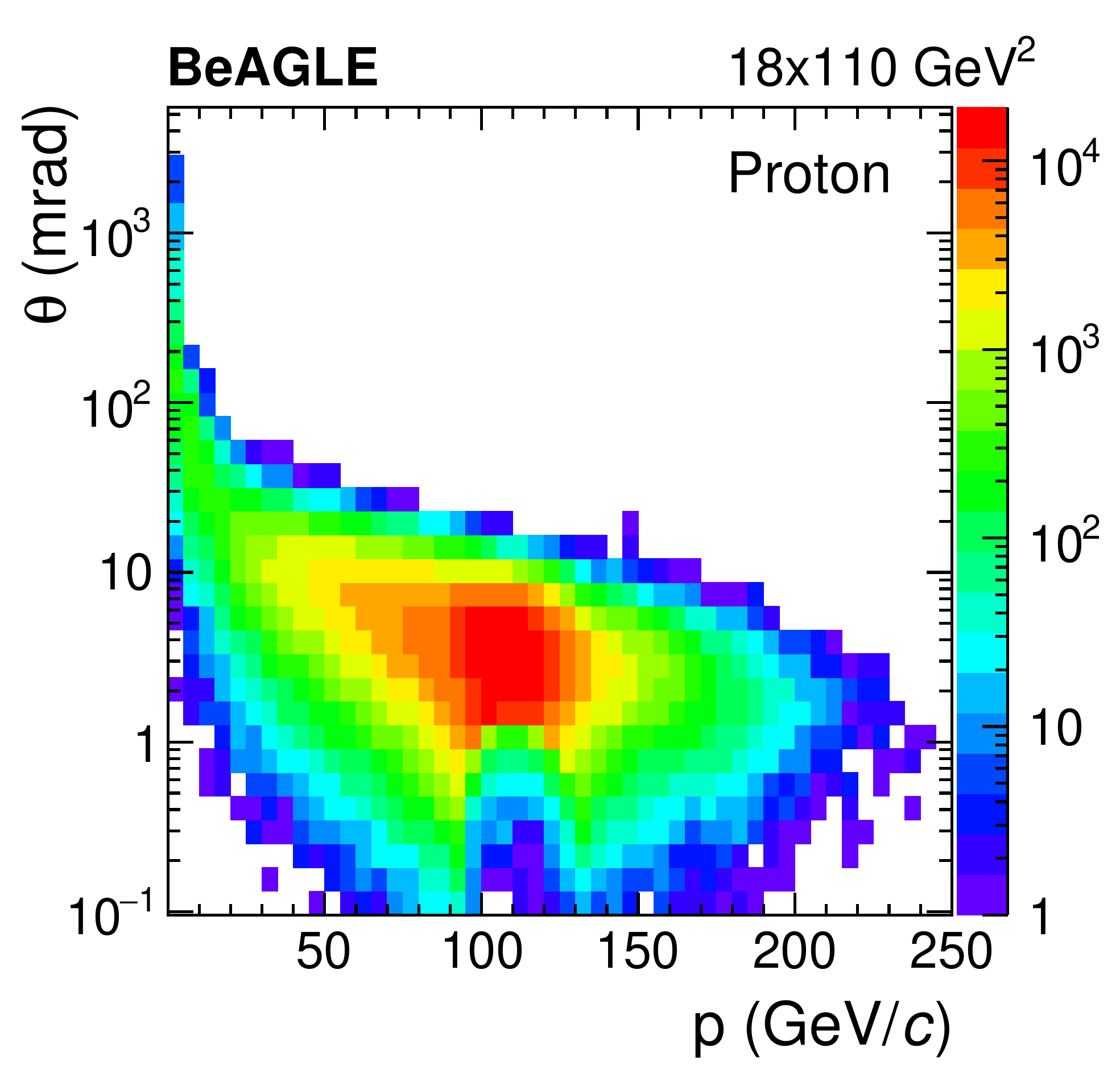}}
\subfigure[]{
\label{fig:figure_4_b}
\includegraphics[width=0.3\textwidth]{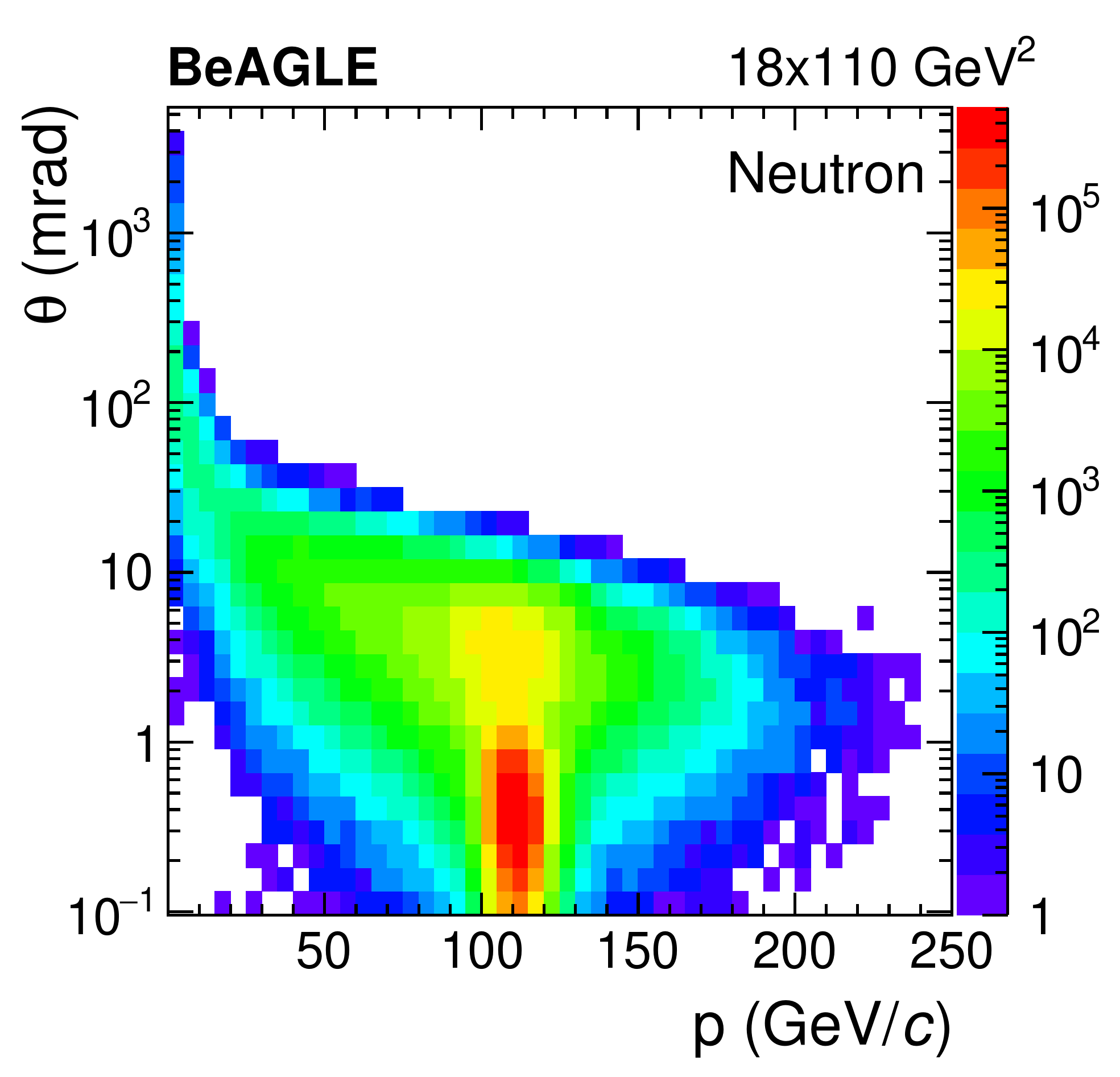}}
\subfigure[]{
\label{fig:figure_4_c}
\includegraphics[width=0.3\textwidth]{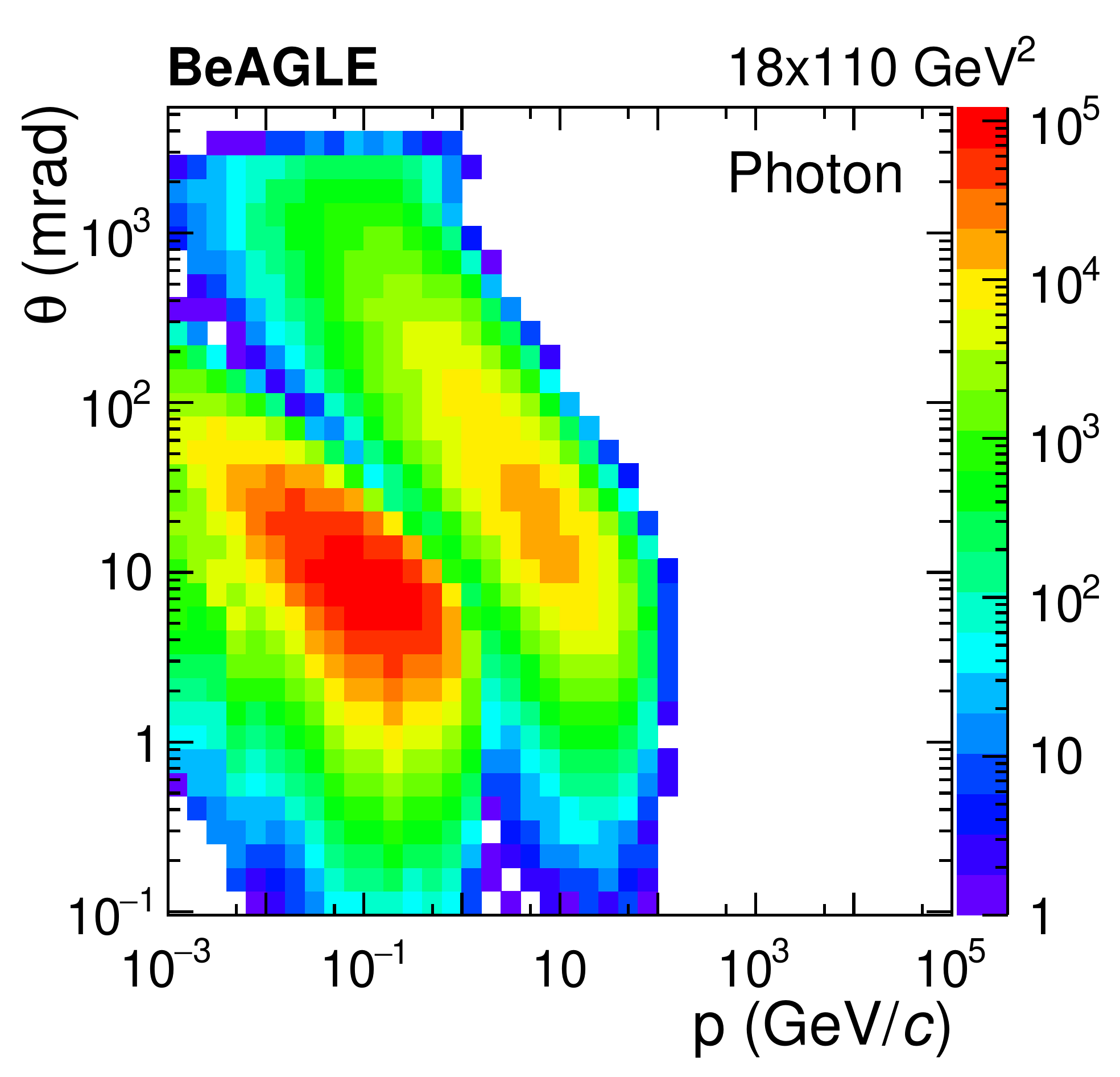}}
\caption{The scattering angle as function of the  total momentum of (a) protons, (b) neutrons, and (c) photons for incoherent events before any veto was applied. }
\label{fig:figure_4}
\end{figure*}

In Fig.~\ref{fig:figure_4}, the two dimensional distributions of scattering angle $\theta$ and the total momentum $p$ of protons, neutrons, and photons are shown for the process of incoherent $J/\psi$ production in $ePb$ collisions with 18 GeV on 110 GeV at the EIC. Based on single particle distributions, neutrons are found to have an acceptance$\times$efficiency close to 75\% within a 5 mrad cone of the scattering angle. Most of the inefficiency is due to the neutrons starting showering already in the beam pipe (see details later). For protons, the acceptance is generally very good except at very large scattering angles and at very low momentum. The three different proton detectors cover almost the entire phase space in the scattering angle up to 22 mrad. Because of the rigidity change, the RPs have an insignificant contribution, while the OMDs and the B0-detectors accept 31\% and 16\% of the protons, respectively. Photons with an energy above 50 MeV and a scattering angle less than 5 mrad can be detected by the ZDC. This results in  a 16\% acceptance. Most of the photons have a scattering angle outside the ZDC acceptance. Therefore, a pre-shower detector is included in the set of detectors to be installed  on B0 magnet bore to detect the photon with a  scattering angle greater than 5 mrad and less than 22 mrad. Note protons and neutrons with very low momentum and very large scattering angle are mostly the primary nucleon participating in the hard scattering or are products of the INC. However, in most cases, these particles are produced in coincidence with other breakup particles within the detector acceptance; therefore, the events can be efficently vetoed. 

For a successful veto of incoherent diffractive events, detection of one particle is sufficient. Therefore, the remaining  background events will have none of the veto requirements fulfilled. In order to detail the vetoing procedure step-by-step, we break them down to selections on different particles in different detectors, listed as follows:

\begin{itemize}
    	\item Veto.1: no activity other than $e^{-}$ and $J/\psi$ in the main detector ( $\left | \eta  \right |<4.0$ and $p_{T} > 100~{\rm MeV}/c $) ; 
    \item Veto.2: Veto.1 and no neutron in ZDC; 
    \item Veto.3: Veto.2 and no proton in RP;
    \item Veto.4: Veto.3 and no proton in OMDs;
    \item Veto.5: Veto.4 and no proton in B0;
    \item Veto.6: Veto.5 and no photon in B0;
    \item Veto.7: Veto.6 and no photon with $E > 50~ \rm{MeV}$ in ZDC.
\end{itemize}

\begin{figure}[tbh]
\centering 
\includegraphics[width=0.4\textwidth]{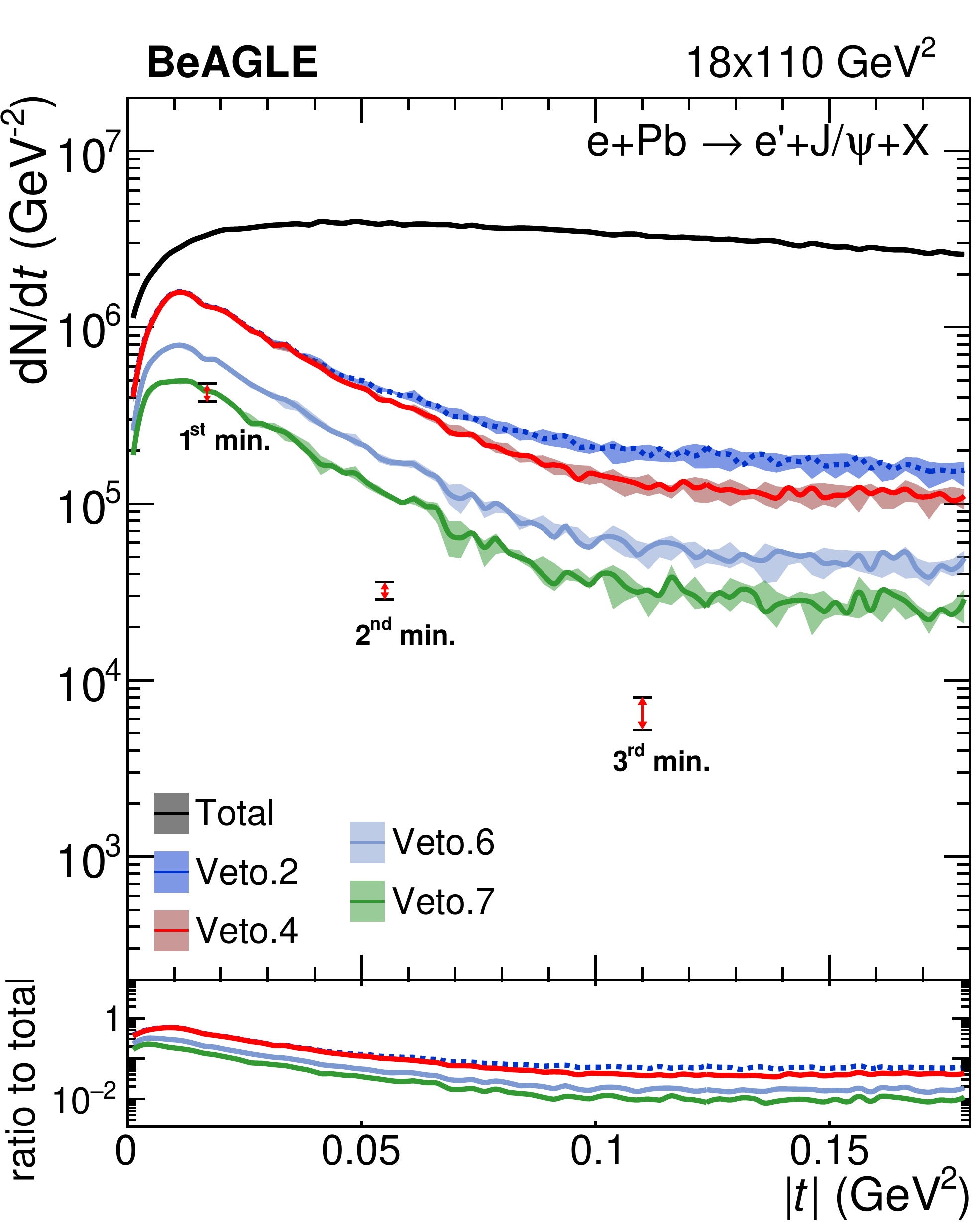}
\caption{Distribution of the momentum transfer $|t|$ for incoherent $J/\psi$ production in $ePb$ collisions with 18 GeV on 110 GeV at the EIC. Different lines indicate results after different vetoing requirements.}  
\label{fig:figure_5}
\end{figure}

\begin{figure*}[th]
\centering 
\subfigure[]{
\label{fig:figure_6_a}
\includegraphics[width=0.3\textwidth]{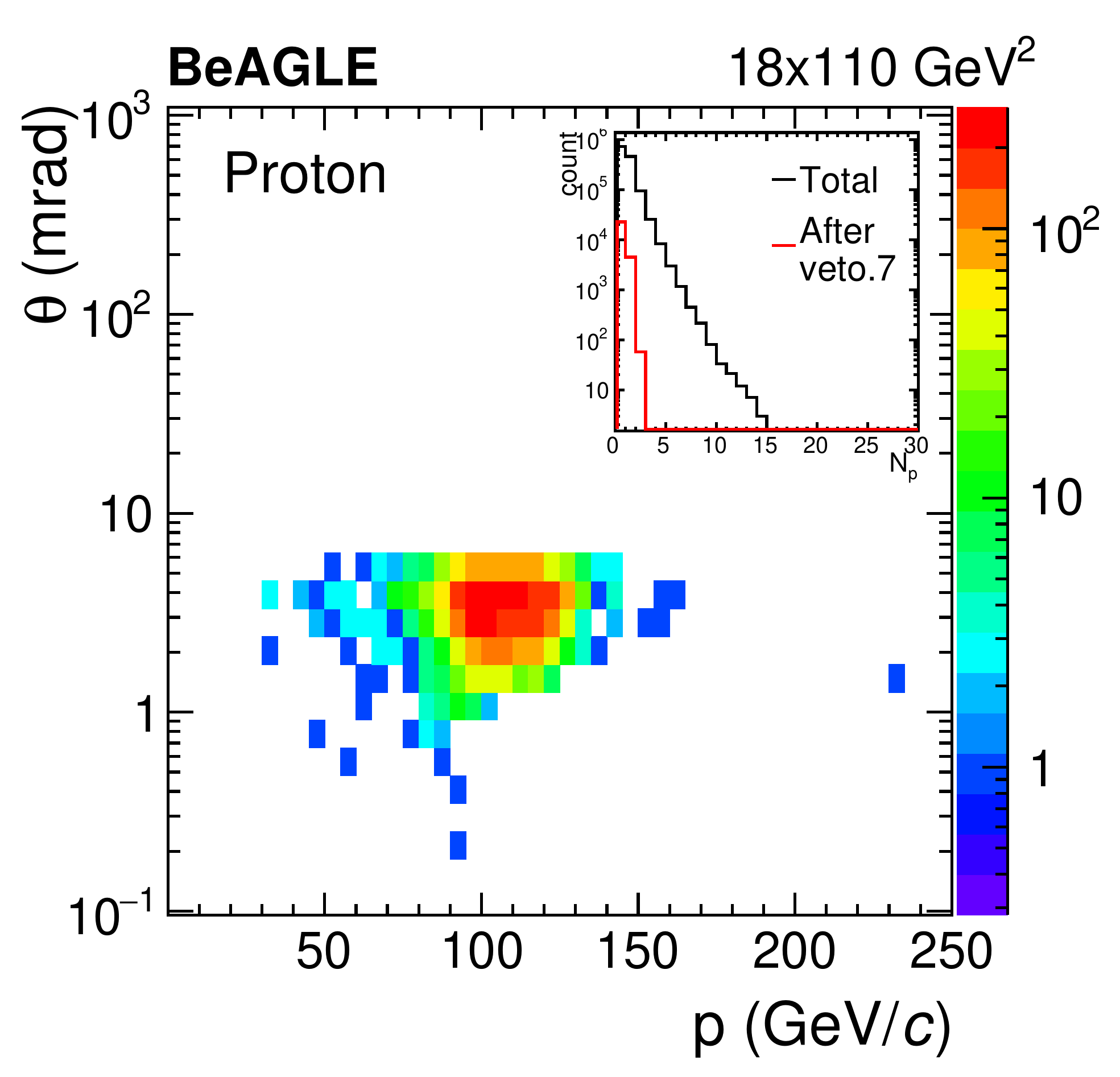}}
\subfigure[]{
\label{fig:figure_6_b}
\includegraphics[width=0.3\textwidth]{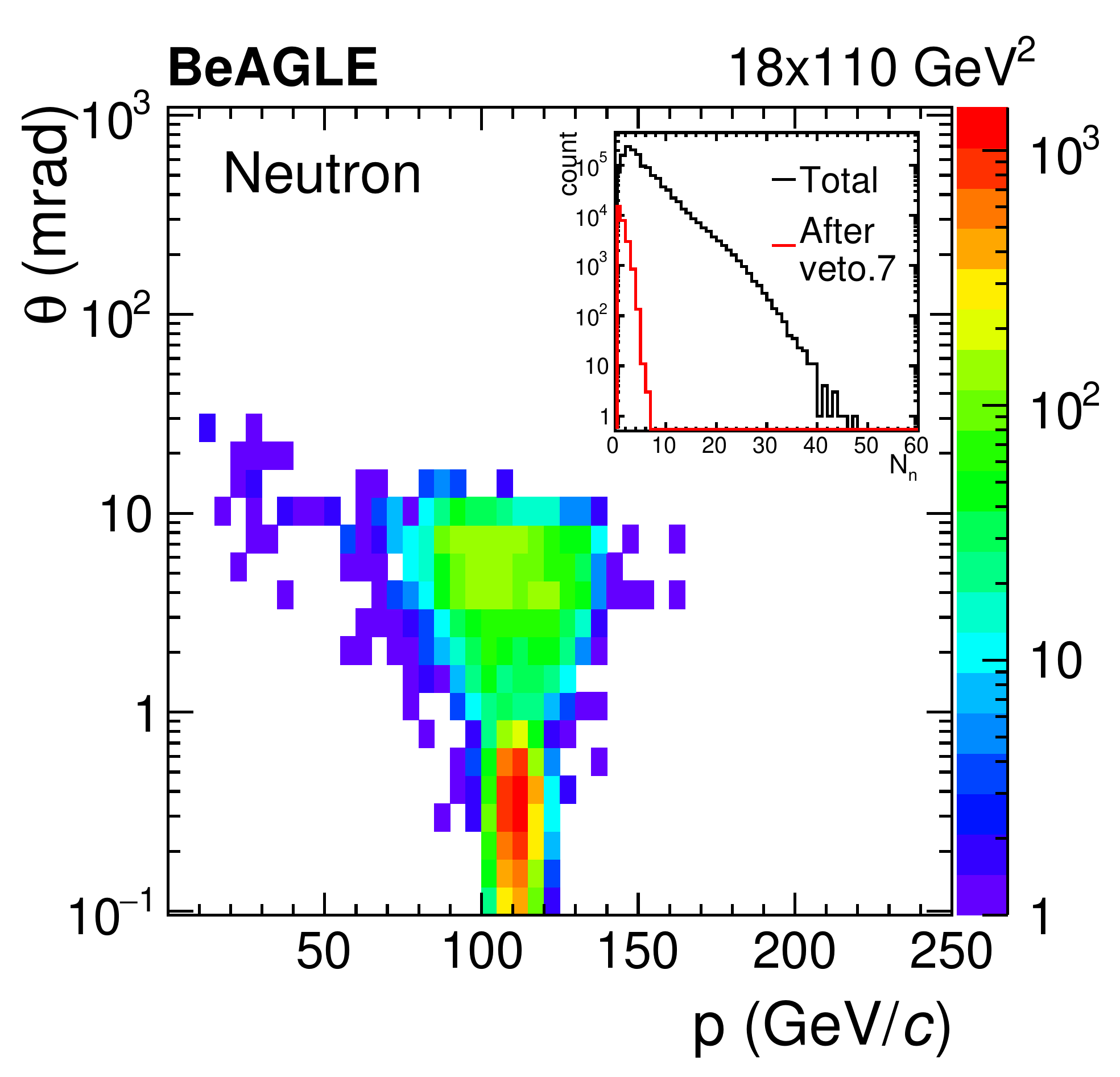}}
\subfigure[]{
\label{fig:figure_6_c}
\includegraphics[width=0.3\textwidth]{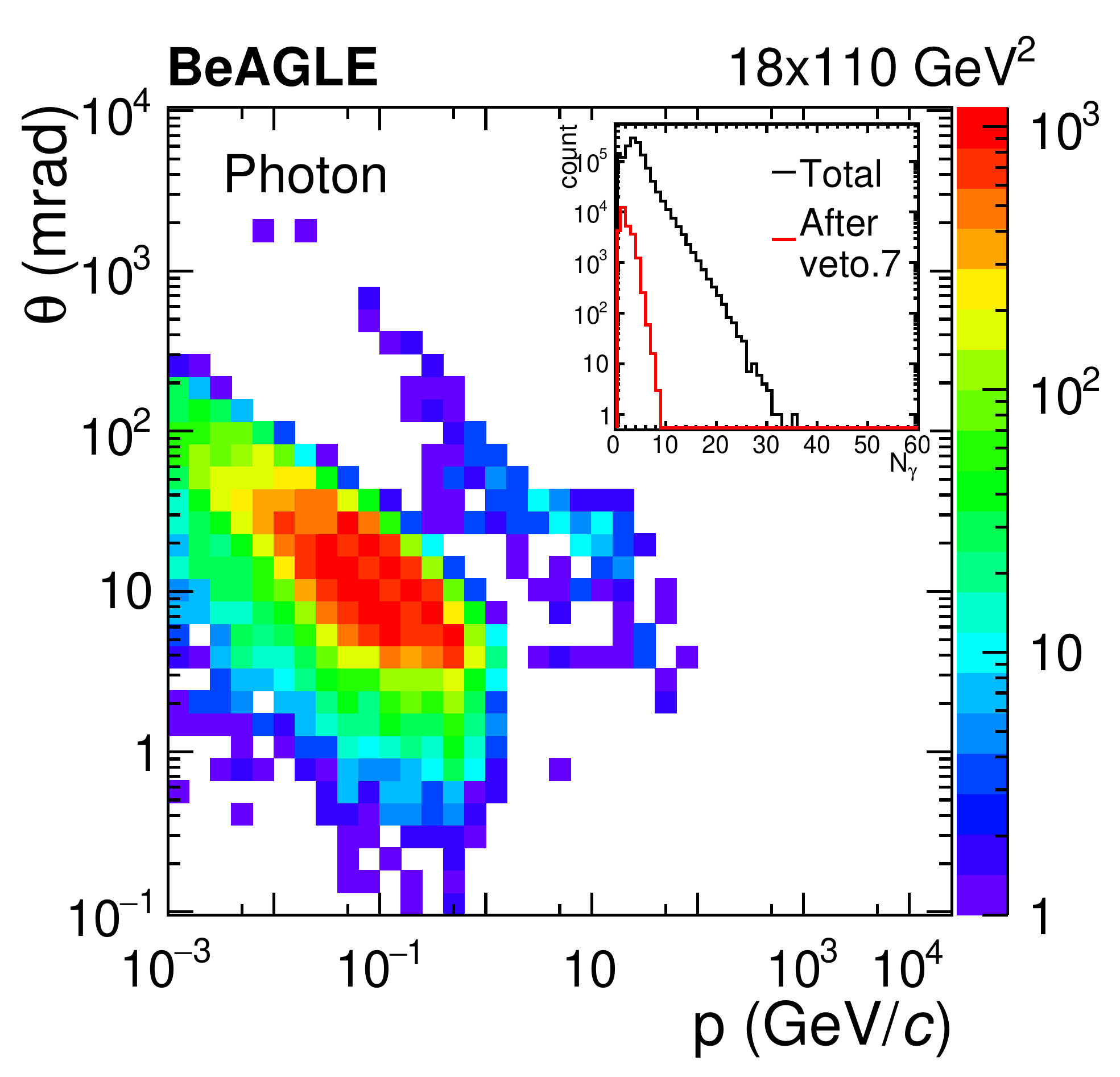}}
\caption{The scattering angle as function of the  total momentum for (a) protons, (b) neutrons, and (c) photons after veto.7. The insert at each panel shows the particle multiplicity distributions, for both the MC truth level (total) and after applying all cuts to veto the incoherent events.}
\label{fig:figure_6}
\end{figure*}

 In Fig.~\ref{fig:figure_5}, the incoherent $J/\psi$ production $dN/d|t|$ as a function of momentum transfer $|t|$ is shown based on the BeAGLE event generator. The total number of events before any vetoing is shown as the black solid line, the other colored lines indicate  the results after different vetoing requirements are applied. The results for cuts  that have negligible impact on the vetoing are not shown in the figure. The uncertainty bands are based on different results obtained by varying the $\tau_{0}$ parameter, from 6~fm to 14~fm with 10~fm used as the central value. The vetoing efficiency with different $\tau_{0}$ values is found to be similar, where the fraction of the total surviving events after veto.7 is 1.98$\%$ for $\tau_{0}=6~$fm, and 2.14$\%$ for $\tau_{0}=14~$fm, respectively. Therefore the quoted uncertainty is less than 0.1$\%$. No detailed studies on the vetoing power of the EIC central detector have been performed as the detector layouts are still being developed. Nevertheless, vetoing particles with $\left | \eta  \right |<4.0$ and $p_{T} > 100~{\rm MeV}/c $ other than the scattered electron $e^{-}$ and $J/\psi$ has a very small impact , because no other events categories, i.e DIS, were  included in this study. Fig.~\ref{fig:figure_5} shows that vetoing on protons, neutrons, and photons are all important and contribute to a significant reduction of the background. After veto.7, the residual contribution is about 1--20\% of the total events, depending on the value of $|t|$. For veto.7, we set a 50 MeV cut as the minimum detection energy of photons. Detecting such low energy photons in the ZDC above background is challenging, therefore we also investigate the effect of a higher cut of 100~MeV. A photon cut of 100~MeV would lead to the fraction of surviving events after veto.7 of about 2.18\% of the total events vs.\ 2.05\% for the default 50~MeV cut, as shown in Table~\ref{tab:table_2} in Appendix~\ref{app:beampipe}.

\begin{figure}[thb]
\centering  
\subfigure[]{
\label{Nucleus.sub.1}
\includegraphics[width=0.4\textwidth]{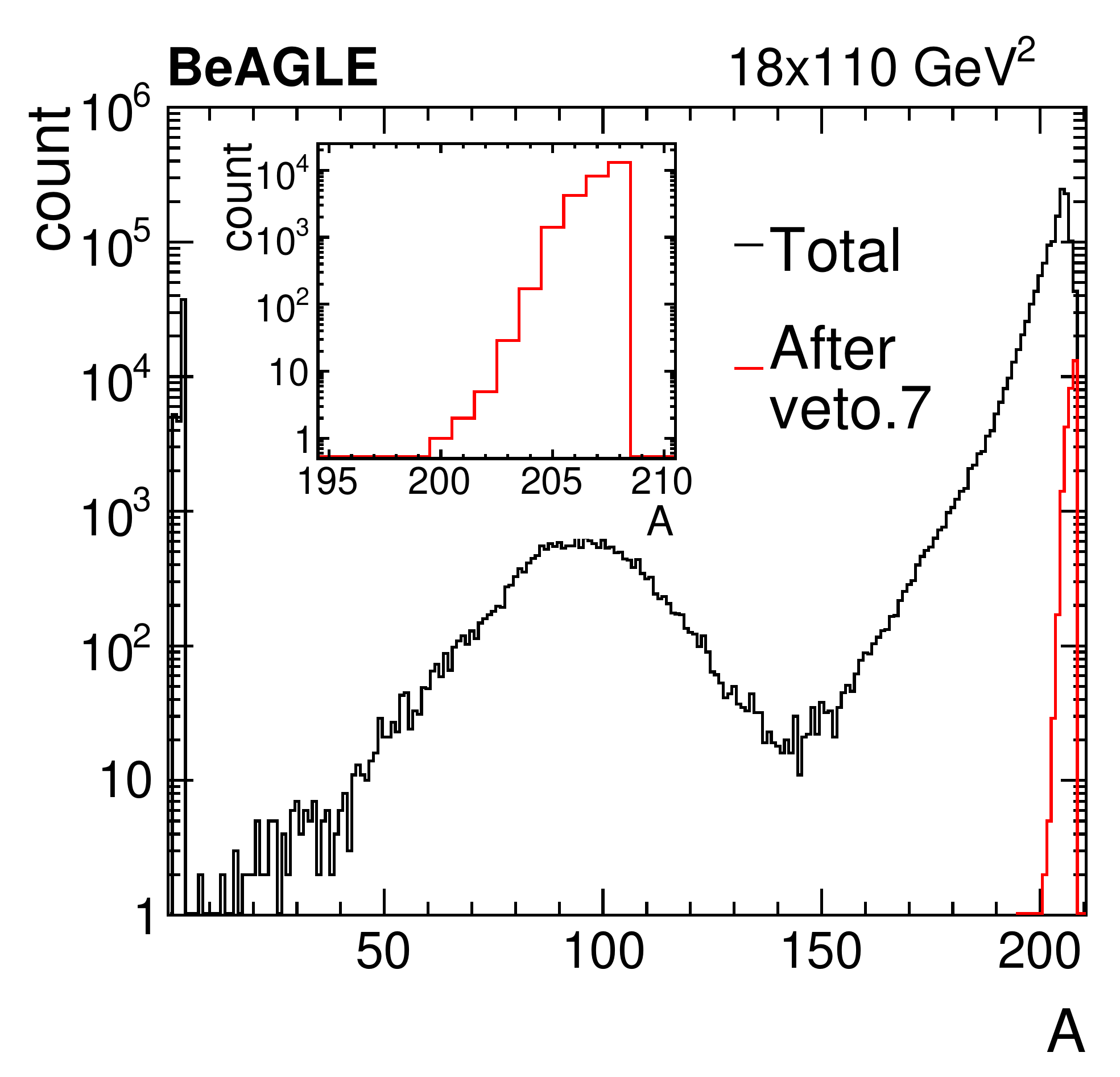}}
\subfigure[]{
\label{Nucleus.sub.2}
\includegraphics[width=0.4\textwidth]{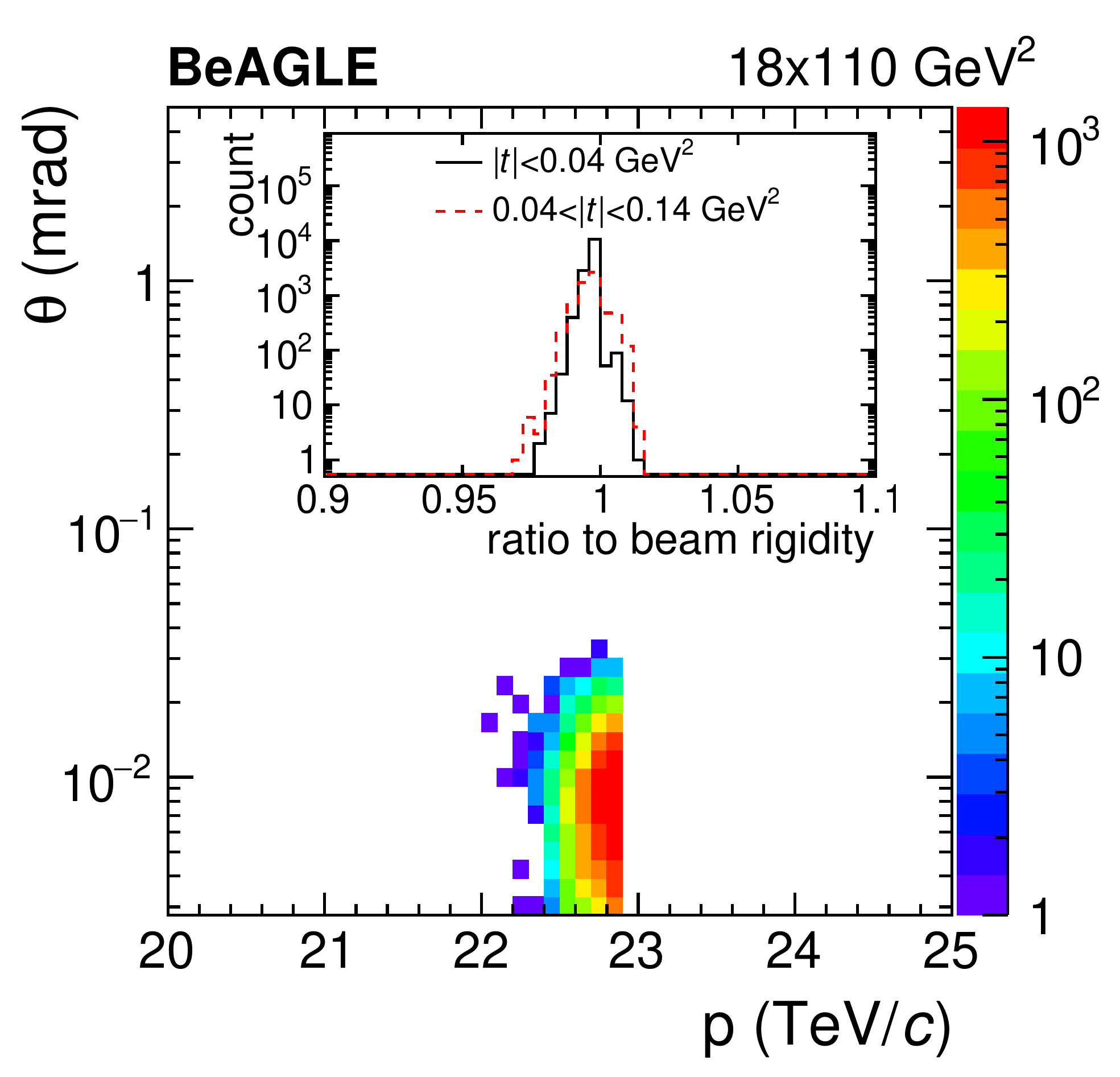}}
\caption{(a) The mass number distribution for nuclei of total and surviving events after veto.7. (b) The scattering angle versus total momentum for nuclei of surviving events after veto.7, the insert shows the ratio of the residual nuclei rigidity to that of the nuclear beam.}
\label{Nucleus_survive}
\end{figure}

\begin{figure}[tbh]
\centering 
\includegraphics[width=0.4\textwidth]{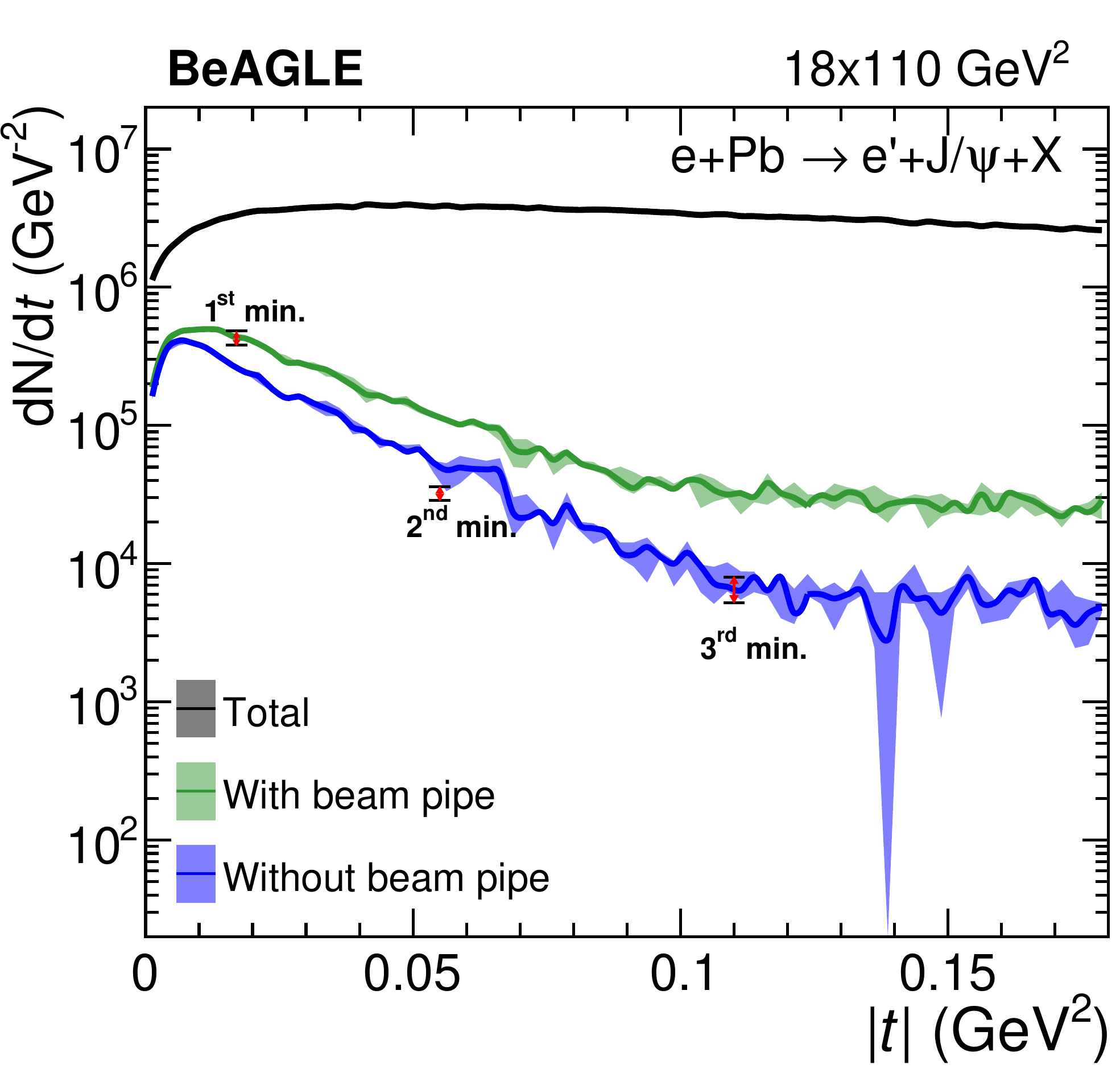}
\caption{Distribution of the momentum transfer $|t|$  for incoherent $J/\psi$ production in $ePb$ collisions at the EIC (18 GeV on 110 GeV). The different lines indicate the effect of the beampipe on the vetoing efficiency .}  
\label{fig:figure_7}
\end{figure} 

Furthermore, the relative magnitude and position of the three coherent diffractive minima based on the Sar\textit{t}re model~\cite{Toll:2012mb} are shown in Fig.~\ref{fig:figure_5} by the red arrows. The difference between the upper and lower bar indicates the difference assuming a saturation and a non-saturation model~\cite{Toll:2012mb}. Even though the result from the Sar\textit{t}re model is for $eAu$ collisions and the BeAGLE study in this paper uses $ePb$ collisions, they are close enough to make this comparison. If one wants to reach the three minima as required to image the parton spatial distribution inside of a nucleus, we find the vetoing efficiency needed is around 99.8\% at the $t$ location of the third minimum. The current result shown in Fig.~\ref{fig:figure_5} is found to be just enough to reach the first minimum, where the needed vetoing efficiency is about 90\%. Based on the Yellow Report~\cite{AbdulKhalek:2021gbh} study, not suppressing the background to the level of these minima, the Fourier transformation to obtain the gluon density distributions would be significantly smeared. So far with the current forward interaction region design and the BeAGLE model, there is at least a factor of 4 or more suppression needed to reach the second and third minima. 

The scattering angle $\theta$ versus total momentum $p$ after veto.7 for protons, neutrons, and photons of the residual events are shown in Fig.~\ref{fig:figure_6}. The insert in each panel show the particle multiplicity distributions, for both before and after vetoing. The vetoing efficiency for neutrons, protons, and photons are more than 99\%. The dominant residual particle among the three is the photon, especially photons with very low energy. The multiplicity of the surviving events is peaked at low multiplicity, because the particles there are in a single event, the more likely it can be vetoed. High multiplicity events are also more likely to have soft particles, which are further away from the nominal beam momentum. These particles are easier to be vetoed; therefore, high multiplicity events are almost always rejected. 

The surviving events not only have low particle multiplicity but also contain  high mass nuclear remnants (e.g., for $Pb$ with A=208, A=207, A=206, etc.) with momenta close to the beam. The mass number distribution for nuclei of total and surviving events after veto.7 are shown in Fig.~\ref{Nucleus.sub.1}. Examining the residual events, the mass number A is greater than 200 and peaks at 208, which is primarily due to events with only photon emission from an excited $Pb^{208}$ nucleus. Figure.~\ref{Nucleus.sub.2} shows the scattering angle as function of the total momentum for nuclei for these residual events, the momentum is large and the scattering angle is less than 0.05 mrad. The rigidity change of these nuclei is presented as the ratio with respect to the beam rigidity, see sub-panel of Fig.~\ref{Nucleus.sub.2}.  The black solid line shows the distribution of events with momentum transfer $|t|$ less than 0.04 $\rm GeV^{2}$ and the red dashed line is for events with $0.04<|t|<0.14~ \rm GeV^{2}$. The ratio for both $|t|$ ranges is between 0.97 to 1.03, which makes it extremely difficult to detect. Based on these numbers, these nuclei will remain in the envelope of the beam at IP6. 

Finally, we have also investigated the impact of the beam pipe on the veto efficiency. The results with and without beam pipe are shown in Fig.~\ref{fig:figure_7}. The green line represents the distribution of the residual events including a beam pipe in the simulation while the blue line shows the result without the beam pipe. As one can see, without the beam pipe, the $t$-averaged veto efficiency is close to $99\%$ in total after veto.7 and found to be enough to reach the first, second, and third minimum of the momentum transfer $|t|$ distribution. We studied also the dependence of the veto efficiency on the beam pipe material. The surviving events after veto.7 with an Aluminum beam pipe are $2.46\%$, which is almost the same as for Beryllium. The veto efficiency with Stainless Steel is found to be significantly lower with $5.58\%$ incoherent events surviving. The current design of the beam pipe is still preliminary, further optimization of the beam pipe material and layout could lead to improvements in the efficiency of detecting the nuclear break-up particles. In short, the beam pipe design is critical to the veto efficiency for the process of interest in this study. For details, see Appendix~\ref{app:beampipe} for the surviving event ratio after each step for different materials, as is shown in Table~\ref{tab:table_2}.

\section{\label{sec:discussions} Discussion}

\subsection{\label{subsec:modeluncertainty} Model Uncertainty}

In the prior section, we described the vetoing process of incoherent $J/\psi$ production modeled in the BeAGLE event generator that is experimentally accessible at the EIC. However, the nuclear breakup mechanism, its associated particle production, and their kinematic distributions in $eA$ collisions based on the BeAGLE model, have not been rigorously validated against data. As of today, there is no data that can be used for validation. In the near future, the data that might provide insights to this process are UPCs at RHIC and at the LHC Run 3. Therefore, there remain some theoretical uncertainties in the BeAGLE model. However, we have systematically checked parameters related to the hadron formation time $\tau$, the results are found to be stable with respect to these model variations. 

There is another source of uncertainty which comes from the Sar\textit{t}re model.
Since the BeAGLE generator only predicts the incoherent  $J/\psi$ production, we adopt the ratio of coherent to incoherent cross section from the Sar\textit{t}re model to define the position and relative magnitude of the three minima shown in Fig.~\ref{fig:figure_5} and Fig.~\ref{fig:figure_7}. Similar to BeAGLE, the prediction of incoherent and coherent cross sections from the Sar\textit{t}re generator may have their own model uncertainties, especially for the incoherent production. From a recent study of the Sar\textit{t}re model used describing the UPC data, the agreement was found to be very good~\cite{Sambasivam:2019gdd}. However, the comparison was for the total cross section level without studying the $|t|$ dependence. A more quantitative comparison with the differential cross section data as a function of $|t|$ would be extremely valuable for understanding uncertainties in more detail. 

The main results presented in Sec.~\ref{sec:result} are for the kinematic range $Q^{2}$ greater than 1. In order to estimate the $Q^{2}$ dependence, we also studied the vetoing efficency for three $Q^{2}$ bins: $1<Q^{2}<2 ~ \rm GeV^{2}$,  $2<Q^{2}<5 ~ \rm GeV^{2}$ and $5<Q^{2}<10 ~ \rm GeV^{2}$. We obtained the ratio of coherent and incoherent cross section for the three diffractive minima from the Sar\textit{t}re model. For each $Q^{2}$ bin, we repeated our vetoing analysis separately. The vetoing power in the different $Q^{2}$ bins is found to be almost identical. This result is expected, as the nuclear breakup is not directly correlated to the hard scattering. 

\subsection{\label{subsec:improvement} Possible Instrumental Improvements}
The results presented in this paper are based on the EIC reference detector and conceptual design of the IP-6 IR configuration detailed in the EIC CDR~\cite{ref:EICCDR}. One goal of this analysis is to give ideas for further optimizing the forward-going particle tagging efficiency and associated veto efficiency for incoherent breakup events from diffractive $J/\psi$ production in $eA$ collisions. Given the current design and layout, one of the most important aspects is the design of the beam pipe and vacuum system. Although a quantitative analysis and careful assessment are needed in order to draw definitive conclusions, our results indicate that an improvement of the vetoing power is possible by further optimizing the beam pipe design. This optimization would include using thin, low density materials in regions where particles exit (e.g. beryllium and aluminum where possible), and  the inclusion of an exit window or windows to increase the incident angles for exiting particles and thereby decrease the effective interaction length. Based on the quantitative impact study in the EIC Yellow Report~\cite{AbdulKhalek:2021gbh}, the suppression of incoherent events to a level close to the three diffractive minima will fulfill the requirement of this measurement, e.g., the reconstruction of the gluon density distribution in the $Pb$ nucleus. Figure~\ref{fig:figure_7} shows the momentum transfer, $|t|$, with and without the beam pipe included in the simulation. While it's not possible to completely remove the impact of the beam pipe on the vetoing efficiency, efforts are underway to minimize the negative impact of the design and optimize various components (e.g. tapers, exit window) where possible. With a sufficiently optimized design, this measurement could still be feasible at IP6.

In addition, in a recent preliminary design study on the second IR (the 8 o'clock position in the RHIC complex: IP-8) at the EIC, a different crossing angle and far-forward layout have been proposed in comparison to the IP-6 configuration. The current preliminary IP-8 concept includes a 35 mrad crossing angle, and the inclusion of a secondary focus further downstream, which enables a Roman Pots system to be installed in a region where the 10$\sigma$ safe distance is much smaller ($\sim$1 mm or less). The secondary focus makes use of the so-called ``point-to-point" focusing mechanism, which enables optimization of acceptance at low $p_{T}$. This enables the reduction of the transverse beam size by an order of magnitude or more compared to standard location of the Roman Pots system at IP-6 by forcing the beta-functions just after the secondary focus to be similar to the $\beta^{*}$ values at the IP. This will lead to a significant improvement in the forward-going particle acceptance in scattering angle and rigidity phase space, allowing some of the particles and therefore events shown in Fig.~\ref{Nucleus.sub.2} to be tagged and vetoed. This proposed capability in the preliminary IP-8 design will specifically improve the tagging efficiency of nuclear fragments from breakup of the heavy nucleus, which are normally well within the 10$\sigma$ beam envelope at the nominal location of the Roman Pots, and therefore un-detectable.

In addition to the improvements for detection of low-angle charged particles and nuclear remnants at IP-8, assuming wider aperture magnets can improve the acceptance of neutrons and photons in the ZDC, potentially up to $\sim$9~mrad in some regions of the azimuthal phase space. Based on the particle distributions of the residual events in Fig.~\ref{fig:figure_6}, the combination of the improved neutral particle acceptance and the ability to tag more of the charged nuclear remnants could improve the vetoing efficiency.

\section{\label{sec:summary} Summary}
We present an investigation of the background in coherent diffractive $J/\psi$ production using the BeAGLE event generator for 18 GeV electrons scattering off 110 GeV lead nuclei at the Electron-Ion Collider (EIC). The BeAGLE simulations provide the dominant physics background to the coherent diffractive $J/\psi$ production - the incoherent events. In the BeAGLE model, in-coherent $J/\psi$ production processes result in forward-going particles, e.g., protons, neutrons, and photons due to the nuclear breakup.  After simulating these events using the most up-to-date design of the EIC interaction region, specifically for the outgoing hadron beam direction and its detectors and a beryllium beam pipe, the total vetoing fraction of these events is found to be 98\%. This rejection power is found to be just enough to suppress the background events to the same level as the signal events at the first minimum position of the predicted diffractive coherent $|t|$ distribution, while more suppression is needed to reach the level of the second and third minimum.
Although an active investigation on other possible instrumental improvements is on-going within the EIC community, the quantitative study reported in this paper shows for the first time a realistic assessment of realizing this experimental measurement. The method and experimental setup employed in this work will serve as a baseline for future design iterations of the EIC forward IR and its detectors.

\begin{acknowledgments}
We thank T. Ullrich for discussion on the Sar\textit{t}re model and its prediction for different kinematic ranges in $Q^{2}$. We thank C.Hyde,  V.~Morozov, T.~Toll and P.~Turonski for helpful discussions.  The authors would also like to thank the EIC project interaction region working group for their help in understanding the various impacts of the IR design. The work of W. Chang is supported by the U.S. Department of Energy under Contract No. de-sc0012704 and the National Natural Science Foundation of China with Grant No. 11875143. The work of E.C. Aschenauer, A. Jentsch, and JH. Lee is supported by the U.S. Department of Energy under Contract No. de-sc0012704, and A. Jentsch is also supported by the Program Development program at Brookhaven National Laboratory. The work of M.~D.~Baker is supported by DOE contracts de-sc0012704, DE-AC05-06OR23177, and by Jefferson Lab LDRD project LDRD1706. The work of Z. Tu is supported by LDRD-039, the U.S. Department of Energy under Award DE-SC0012704, and the Goldhaber Distinguished Fellowship at Brookhaven National Laboratory. The work of Z. Yin is supported by the National Natural Science Foundation of China with Grant No. 11875143. The work of L. Zheng is supported by National Natural Science Foundation of China under Grant No. 11905188. 
\end{acknowledgments}
\appendix

\section{Determination of the $\tau_{0}$ parameter in BeAGLE}
\label{app:tau0}

In BeAGLE (and in DPMJET), a formation time $\tau$ is needed before the secondary particles can be newly created and interact with other nucleons. It is defined as Eq.~(\ref{equation:formationtime}) in Sec.~\ref{sec:generator}. $\tau_{0}$ is a free parameter which determines the overall scale for the formation length in this equation, and its setting can affect the multiplicity of neutron emission. 

\begin{figure}[thb]
\centering
\includegraphics[width=3.3in]{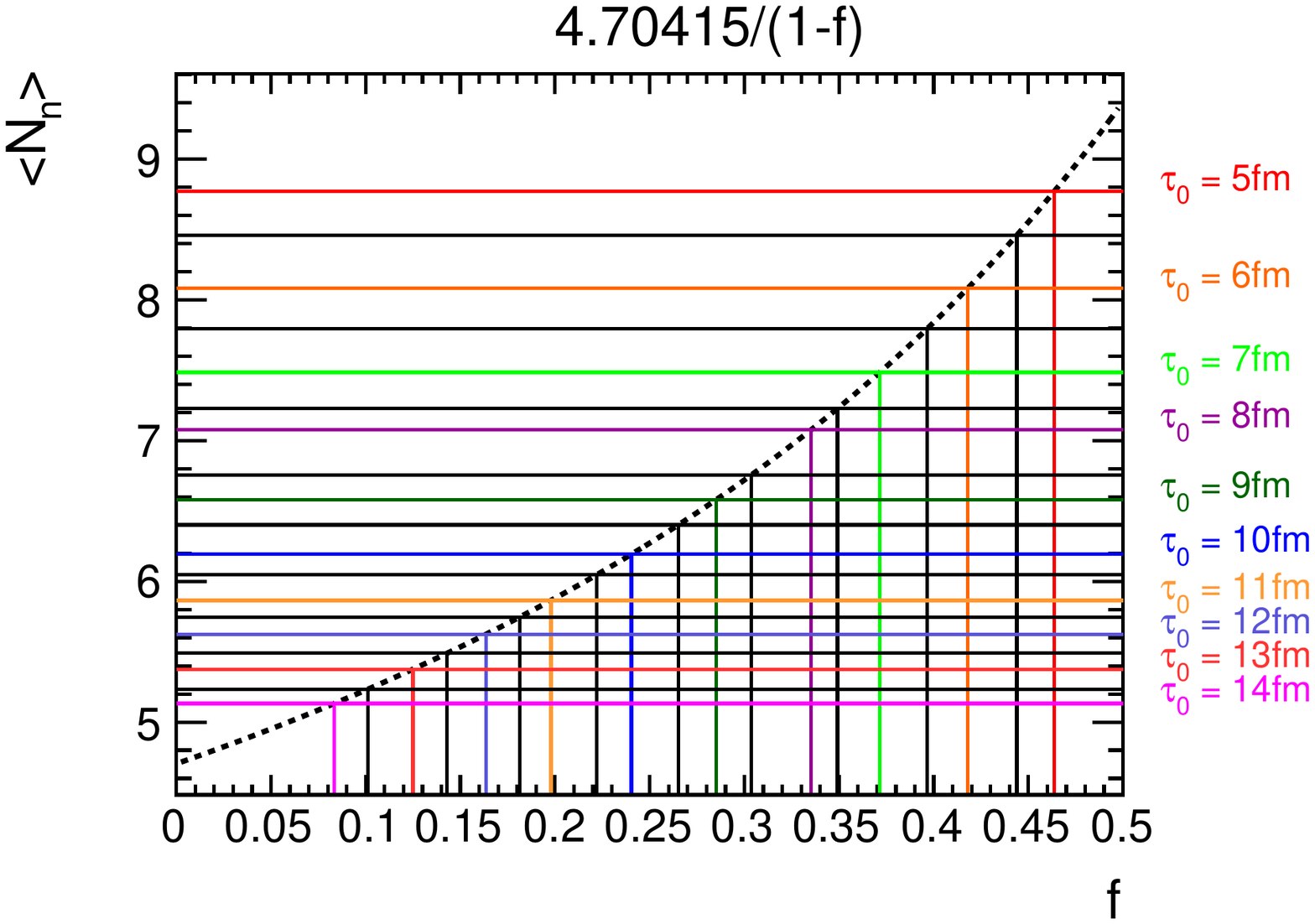}
  \caption{ \label{fig:tau0}  The average neutron multiplicity $\left \langle N_{n} \right \rangle$, vs.\ $f$, the ratio of the number of coherent to the total events for different values of $\tau_0$.}
\end{figure}

 \begin{table*}[thb]
 \fontsize{11}{13}\selectfont
\centering
\begin{tabular}{c|c|c|c|c}
\hline
\multicolumn{5}{c}{Survived Event Ratio}\\ \hline 
Material &  Without beam pipe & Beryllium & Aluminum & Stainless Steel\\\hline
Total events & 100 \% & 100 \%  & 100\% & 100\%\\\hline
Veto.1 & 86.9\%  & 86.9\% & 86.9\% & 86.9 \%\\\hline
Veto.2 & 5.81\% & 9.73\% & 9.85\% & 17.2\%  \\\hline
Veto.3 & 5.81\% & 9.73 \% & 9.85\% & 17.2\% \\\hline
Veto.4 & 5.09\% & 8.77\%  & 8.89\% & 15.73\% \\\hline
Veto.5 & 4.32\% & 6.22\%  & 5.97\% & 10.18\% \\\hline
Veto.6 & 2.29\% & 3.32\% & 3.18\% & 5.68\%  \\\hline
Veto.7 ($E_{\rm photon} > 50~ \rm{MeV}$) & 1.06\% &2.05\% & 2.46\% & 5.58\% \\\hline
Veto.7 ($E_{\rm photon} > 100~ \rm{MeV}$) & - & 2.18\% & - & - \\\hline
\end{tabular}
\caption{Summary of the percentage of events surviving the different vetoing steps for incoherent events assuming no beam pipe and different beam pipe materials of beryllium, aluminum, and stainless steel.}
\label{tab:table_2}
\end{table*}

Fig.~\ref{fig:tau0} shows the average neutron multiplicity, $\left \langle N_{n} \right \rangle$ vs.\ the fraction of coherent to total events, $f$, for a variety of different $\tau_{0}$ values. The black dashed curve in the plot is constrained by E665 neutron data, the value of 4.70415 is obtained by a fit to E665 neutron multiplicity as a function of $\nu$ for $Pb$ target \cite{E665:1995utr}. After fitting, we found $N_{n}({\rm E665})$ is independent of $\nu$, which is a constant of 4.70415. From Eq.~(\ref{equation:E665}), one can get that $N_{n}({\rm BeAGLE}) = 4.70415 /(1-f)$. Different colors represent different $\tau_{0}$ values, where the black solid lines are the half-integer $\tau_{0}$ values. For example, the one between 5~fm and 6~fm are for $\tau_{0}$ = 5.5~fm, similar for others.  We use the following settings for the main results: $\tau_{0}=6$~fm ($f=0.42$), 10~fm ($f=0.24$) and 14~fm ($f=0.08$). The result  with $ \tau_{0}$= 10 fm is used as the central value, while 6~fm and 14~fm represent the systematic uncertainty. 

\section{Beam pipe material study}
\label{app:beampipe}

In the present study, a simplified beam pipe design was included in the simulations to quantify the impact on the veto efficiency. A final engineering design was not available at the time of the study for consideration. The beam pipe sections will be briefly in order from the IP to the drift region where the far-forward detectors (RP, OMD, and ZDC) are located. The beam pipe in the B0pf dipole magnet has a radius of 3.2~cm and is made of beryllium with a thickness of 2~mm. This radius was chosen to allow for a 5~mrad cone of protons to pass though unimpeded. The beam pipe sections in the other dipoles and quadrupoles between B0pf and B1apf have the same radii as their respective magnet bores, and assume stainless tubes with a 2~mm thickness. The beam pipe in the drift region after the B1apf dipole magnet consists of two sections spanning the full length of the drift region: i) a straight cylindrical tube with a radius of 20~cm that extends just past the second Roman Pots station, ii) a conical section that tapers from a 20~cm radius to a 7~cm radius at the B2pf magnet entrance.
Both sections of the beam pipe between B1apf and B2pf are made of stainless steel with a thickness of 2~mm. However, simulations were also performed using beryllium and aluminum since sections of the pipe where protons, neutrons, and photons exit to stream toward detectors could consist of less dense material sections or modest exit windows, or both. The final design considerations are still under study.  

Table~\ref{tab:table_2} summarizes the step-by-step vetoing power assuming different beam pipe material. Note the geometrical layout is identical in all cases. For veto.7, the percentage of survived events with a higher energy cut of 100 MeV for detecting photons is also shown in the bottom row for the beryllium case. This systematic variation is to ensure that our conclusion drawn on the incoherent vetoing is not sensitive to the energy threshold of the photon detection in these kinematic ranges.

\bibliography{beagle}

\begin{thebibliography}{38}%
\makeatletter
\providecommand \@ifxundefined [1]{%
 \@ifx{#1\undefined}
}%
\providecommand \@ifnum [1]{%
 \ifnum #1\expandafter \@firstoftwo
 \else \expandafter \@secondoftwo
 \fi
}%
\providecommand \@ifx [1]{%
 \ifx #1\expandafter \@firstoftwo
 \else \expandafter \@secondoftwo
 \fi
}%
\providecommand \natexlab [1]{#1}%
\providecommand \enquote  [1]{``#1''}%
\providecommand \bibnamefont  [1]{#1}%
\providecommand \bibfnamefont [1]{#1}%
\providecommand \citenamefont [1]{#1}%
\providecommand \href@noop [0]{\@secondoftwo}%
\providecommand \href [0]{\begingroup \@sanitize@url \@href}%
\providecommand \@href[1]{\@@startlink{#1}\@@href}%
\providecommand \@@href[1]{\endgroup#1\@@endlink}%
\providecommand \@sanitize@url [0]{\catcode `\\12\catcode `\$12\catcode
  `\&12\catcode `\#12\catcode `\^12\catcode `\_12\catcode `\%12\relax}%
\providecommand \@@startlink[1]{}%
\providecommand \@@endlink[0]{}%
\providecommand \url  [0]{\begingroup\@sanitize@url \@url }%
\providecommand \@url [1]{\endgroup\@href {#1}{\urlprefix }}%
\providecommand \urlprefix  [0]{URL }%
\providecommand \Eprint [0]{\href }%
\providecommand \doibase [0]{http://dx.doi.org/}%
\providecommand \selectlanguage [0]{\@gobble}%
\providecommand \bibinfo  [0]{\@secondoftwo}%
\providecommand \bibfield  [0]{\@secondoftwo}%
\providecommand \translation [1]{[#1]}%
\providecommand \BibitemOpen [0]{}%
\providecommand \bibitemStop [0]{}%
\providecommand \bibitemNoStop [0]{.\EOS\space}%
\providecommand \EOS [0]{\spacefactor3000\relax}%
\providecommand \BibitemShut  [1]{\csname bibitem#1\endcsname}%
\let\auto@bib@innerbib\@empty
\bibitem [{\citenamefont {Adam}\ \emph {et~al.}(2021)\citenamefont {Adam} \emph
  {et~al.}}]{ref:EICCDR}%
  \BibitemOpen
  \bibfield  {author} {\bibinfo {author} {\bibfnamefont {J.}~\bibnamefont
  {Adam}} \emph {et~al.},\ }\href@noop {} {\enquote {\bibinfo {title} {Electron
  ion collider conceptual design report},}\ }\bibinfo {howpublished}
  {\url{https://www.bnl.gov/ec/files/EIC_CDR_Final.pdf}} (\bibinfo {year}
  {2021})\BibitemShut {NoStop}%
\bibitem [{\citenamefont {Accardi}\ \emph {et~al.}(2016)\citenamefont {Accardi}
  \emph {et~al.}}]{Accardi:2012qut}%
  \BibitemOpen
  \bibfield  {author} {\bibinfo {author} {\bibfnamefont {A.}~\bibnamefont
  {Accardi}} \emph {et~al.},\ }\href {\doibase 10.1140/epja/i2016-16268-9}
  {\bibfield  {journal} {\bibinfo  {journal} {Eur. Phys. J. A}\ }\textbf
  {\bibinfo {volume} {52}},\ \bibinfo {pages} {268} (\bibinfo {year} {2016})},\
  \Eprint {http://arxiv.org/abs/1212.1701} {arXiv:1212.1701 [nucl-ex]}
  \BibitemShut {NoStop}%
\bibitem [{\citenamefont {Munier}\ \emph {et~al.}(2001)\citenamefont {Munier},
  \citenamefont {Stasto},\ and\ \citenamefont {Mueller}}]{Munier:2001nr}%
  \BibitemOpen
  \bibfield  {author} {\bibinfo {author} {\bibfnamefont {S.}~\bibnamefont
  {Munier}}, \bibinfo {author} {\bibfnamefont {A.~M.}\ \bibnamefont {Stasto}},
  \ and\ \bibinfo {author} {\bibfnamefont {A.~H.}\ \bibnamefont {Mueller}},\
  }\href {\doibase 10.1016/S0550-3213(01)00168-7} {\bibfield  {journal}
  {\bibinfo  {journal} {Nucl. Phys. B}\ }\textbf {\bibinfo {volume} {603}},\
  \bibinfo {pages} {427} (\bibinfo {year} {2001})},\ \Eprint
  {http://arxiv.org/abs/hep-ph/0102291} {arXiv:hep-ph/0102291} \BibitemShut
  {NoStop}%
\bibitem [{\citenamefont {Gelis}\ \emph {et~al.}(2010)\citenamefont {Gelis},
  \citenamefont {Iancu}, \citenamefont {Jalilian-Marian},\ and\ \citenamefont
  {Venugopalan}}]{Gelis:2010nm}%
  \BibitemOpen
  \bibfield  {author} {\bibinfo {author} {\bibfnamefont {F.}~\bibnamefont
  {Gelis}}, \bibinfo {author} {\bibfnamefont {E.}~\bibnamefont {Iancu}},
  \bibinfo {author} {\bibfnamefont {J.}~\bibnamefont {Jalilian-Marian}}, \ and\
  \bibinfo {author} {\bibfnamefont {R.}~\bibnamefont {Venugopalan}},\ }\href
  {\doibase 10.1146/annurev.nucl.010909.083629} {\bibfield  {journal} {\bibinfo
   {journal} {Ann. Rev. Nucl. Part. Sci.}\ }\textbf {\bibinfo {volume} {60}},\
  \bibinfo {pages} {463} (\bibinfo {year} {2010})},\ \Eprint
  {http://arxiv.org/abs/1002.0333} {arXiv:1002.0333 [hep-ph]} \BibitemShut
  {NoStop}%
\bibitem [{\citenamefont {Jalilian-Marian}(2014)}]{Jalilian-Marian:2014ica}%
  \BibitemOpen
  \bibfield  {author} {\bibinfo {author} {\bibfnamefont {J.}~\bibnamefont
  {Jalilian-Marian}},\ }\href {\doibase 10.1051/epjconf/20146604012} {\bibfield
   {journal} {\bibinfo  {journal} {EPJ Web Conf.}\ }\textbf {\bibinfo {volume}
  {66}},\ \bibinfo {pages} {04012} (\bibinfo {year} {2014})}\BibitemShut
  {NoStop}%
\bibitem [{\citenamefont {Jalilian-Marian}\ and\ \citenamefont
  {Kovchegov}(2006)}]{Jalilian-Marian:2005ccm}%
  \BibitemOpen
  \bibfield  {author} {\bibinfo {author} {\bibfnamefont {J.}~\bibnamefont
  {Jalilian-Marian}}\ and\ \bibinfo {author} {\bibfnamefont {Y.~V.}\
  \bibnamefont {Kovchegov}},\ }\href {\doibase 10.1016/j.ppnp.2005.07.002}
  {\bibfield  {journal} {\bibinfo  {journal} {Prog. Part. Nucl. Phys.}\
  }\textbf {\bibinfo {volume} {56}},\ \bibinfo {pages} {104} (\bibinfo {year}
  {2006})},\ \Eprint {http://arxiv.org/abs/hep-ph/0505052}
  {arXiv:hep-ph/0505052} \BibitemShut {NoStop}%
\bibitem [{\citenamefont {Weigert}(2005)}]{Weigert:2005us}%
  \BibitemOpen
  \bibfield  {author} {\bibinfo {author} {\bibfnamefont {H.}~\bibnamefont
  {Weigert}},\ }\href {\doibase 10.1016/j.ppnp.2005.01.029} {\bibfield
  {journal} {\bibinfo  {journal} {Prog. Part. Nucl. Phys.}\ }\textbf {\bibinfo
  {volume} {55}},\ \bibinfo {pages} {461} (\bibinfo {year} {2005})},\ \Eprint
  {http://arxiv.org/abs/hep-ph/0501087} {arXiv:hep-ph/0501087} \BibitemShut
  {NoStop}%
\bibitem [{\citenamefont {Iancu}\ and\ \citenamefont
  {Venugopalan}(2003)}]{Iancu:2003xm}%
  \BibitemOpen
  \bibfield  {author} {\bibinfo {author} {\bibfnamefont {E.}~\bibnamefont
  {Iancu}}\ and\ \bibinfo {author} {\bibfnamefont {R.}~\bibnamefont
  {Venugopalan}},\ }\enquote {\bibinfo {title} {{The Color glass condensate and
  high-energy scattering in QCD}},}\ in\ \href {\doibase
  10.1142/9789812795533_0005} {\emph {\bibinfo {booktitle} {{Quark-gluon plasma
  4}}}},\ \bibinfo {editor} {edited by\ \bibinfo {editor} {\bibfnamefont
  {R.~C.}\ \bibnamefont {Hwa}}\ and\ \bibinfo {editor} {\bibfnamefont {X.-N.}\
  \bibnamefont {Wang}}}\ (\bibinfo {year} {2003})\ \Eprint
  {http://arxiv.org/abs/hep-ph/0303204} {arXiv:hep-ph/0303204} \BibitemShut
  {NoStop}%
\bibitem [{\citenamefont {Kowalski}\ and\ \citenamefont
  {Teaney}(2003)}]{Kowalski:2003hm}%
  \BibitemOpen
  \bibfield  {author} {\bibinfo {author} {\bibfnamefont {H.}~\bibnamefont
  {Kowalski}}\ and\ \bibinfo {author} {\bibfnamefont {D.}~\bibnamefont
  {Teaney}},\ }\href {\doibase 10.1103/PhysRevD.68.114005} {\bibfield
  {journal} {\bibinfo  {journal} {Phys. Rev. D}\ }\textbf {\bibinfo {volume}
  {68}},\ \bibinfo {pages} {114005} (\bibinfo {year} {2003})},\ \Eprint
  {http://arxiv.org/abs/hep-ph/0304189} {arXiv:hep-ph/0304189} \BibitemShut
  {NoStop}%
\bibitem [{\citenamefont {Toll}\ and\ \citenamefont
  {Ullrich}(2013)}]{Toll:2012mb}%
  \BibitemOpen
  \bibfield  {author} {\bibinfo {author} {\bibfnamefont {T.}~\bibnamefont
  {Toll}}\ and\ \bibinfo {author} {\bibfnamefont {T.}~\bibnamefont {Ullrich}},\
  }\href {\doibase 10.1103/PhysRevC.87.024913} {\bibfield  {journal} {\bibinfo
  {journal} {Phys. Rev. C}\ }\textbf {\bibinfo {volume} {87}},\ \bibinfo
  {pages} {024913} (\bibinfo {year} {2013})},\ \Eprint
  {http://arxiv.org/abs/1211.3048} {arXiv:1211.3048 [hep-ph]} \BibitemShut
  {NoStop}%
\bibitem [{\citenamefont {Sambasivam}\ \emph {et~al.}(2020)\citenamefont
  {Sambasivam}, \citenamefont {Toll},\ and\ \citenamefont
  {Ullrich}}]{Sambasivam:2019gdd}%
  \BibitemOpen
  \bibfield  {author} {\bibinfo {author} {\bibfnamefont {B.}~\bibnamefont
  {Sambasivam}}, \bibinfo {author} {\bibfnamefont {T.}~\bibnamefont {Toll}}, \
  and\ \bibinfo {author} {\bibfnamefont {T.}~\bibnamefont {Ullrich}},\ }\href
  {\doibase 10.1016/j.physletb.2020.135277} {\bibfield  {journal} {\bibinfo
  {journal} {Phys. Lett. B}\ }\textbf {\bibinfo {volume} {803}},\ \bibinfo
  {pages} {135277} (\bibinfo {year} {2020})},\ \Eprint
  {http://arxiv.org/abs/1910.02899} {arXiv:1910.02899 [hep-ph]} \BibitemShut
  {NoStop}%
\bibitem [{\citenamefont {Good}\ and\ \citenamefont
  {Walker}(1960)}]{PhysRev.120.1857}%
  \BibitemOpen
  \bibfield  {author} {\bibinfo {author} {\bibfnamefont {M.~L.}\ \bibnamefont
  {Good}}\ and\ \bibinfo {author} {\bibfnamefont {W.~D.}\ \bibnamefont
  {Walker}},\ }\href {\doibase 10.1103/PhysRev.120.1857} {\bibfield  {journal}
  {\bibinfo  {journal} {Phys. Rev.}\ }\textbf {\bibinfo {volume} {120}},\
  \bibinfo {pages} {1857} (\bibinfo {year} {1960})}\BibitemShut {NoStop}%
\bibitem [{\citenamefont {Klein}\ \emph {et~al.}(2017)\citenamefont {Klein},
  \citenamefont {Nystrand}, \citenamefont {Seger}, \citenamefont {Gorbunov},\
  and\ \citenamefont {Butterworth}}]{Klein:2016yzr}%
  \BibitemOpen
  \bibfield  {author} {\bibinfo {author} {\bibfnamefont {S.~R.}\ \bibnamefont
  {Klein}}, \bibinfo {author} {\bibfnamefont {J.}~\bibnamefont {Nystrand}},
  \bibinfo {author} {\bibfnamefont {J.}~\bibnamefont {Seger}}, \bibinfo
  {author} {\bibfnamefont {Y.}~\bibnamefont {Gorbunov}}, \ and\ \bibinfo
  {author} {\bibfnamefont {J.}~\bibnamefont {Butterworth}},\ }\href {\doibase
  10.1016/j.cpc.2016.10.016} {\bibfield  {journal} {\bibinfo  {journal}
  {Comput. Phys. Commun.}\ }\textbf {\bibinfo {volume} {212}},\ \bibinfo
  {pages} {258} (\bibinfo {year} {2017})},\ \Eprint
  {http://arxiv.org/abs/1607.03838} {arXiv:1607.03838 [hep-ph]} \BibitemShut
  {NoStop}%
\bibitem [{\citenamefont {Toll}\ and\ \citenamefont
  {Ullrich}(2014)}]{Toll:2013gda}%
  \BibitemOpen
  \bibfield  {author} {\bibinfo {author} {\bibfnamefont {T.}~\bibnamefont
  {Toll}}\ and\ \bibinfo {author} {\bibfnamefont {T.}~\bibnamefont {Ullrich}},\
  }\href {\doibase 10.1016/j.cpc.2014.03.010} {\bibfield  {journal} {\bibinfo
  {journal} {Comput. Phys. Commun.}\ }\textbf {\bibinfo {volume} {185}},\
  \bibinfo {pages} {1835} (\bibinfo {year} {2014})},\ \Eprint
  {http://arxiv.org/abs/1307.8059} {arXiv:1307.8059 [hep-ph]} \BibitemShut
  {NoStop}%
\bibitem [{\citenamefont {Abdul~Khalek}\ \emph {et~al.}(2021)\citenamefont
  {Abdul~Khalek} \emph {et~al.}}]{AbdulKhalek:2021gbh}%
  \BibitemOpen
  \bibfield  {author} {\bibinfo {author} {\bibfnamefont {R.}~\bibnamefont
  {Abdul~Khalek}} \emph {et~al.},\ }\href@noop {} {\  (\bibinfo {year}
  {2021})},\ \Eprint {http://arxiv.org/abs/2103.05419} {arXiv:2103.05419
  [physics.ins-det]} \BibitemShut {NoStop}%
\bibitem [{\citenamefont {Khachatryan}\ \emph {et~al.}(2017)\citenamefont
  {Khachatryan} \emph {et~al.}}]{Khachatryan:2016qhq}%
  \BibitemOpen
  \bibfield  {author} {\bibinfo {author} {\bibfnamefont {V.}~\bibnamefont
  {Khachatryan}} \emph {et~al.} (\bibinfo {collaboration} {CMS}),\ }\href
  {\doibase 10.1016/j.physletb.2017.07.001} {\bibfield  {journal} {\bibinfo
  {journal} {Phys. Lett. B}\ }\textbf {\bibinfo {volume} {772}},\ \bibinfo
  {pages} {489} (\bibinfo {year} {2017})},\ \Eprint
  {http://arxiv.org/abs/1605.06966} {arXiv:1605.06966 [nucl-ex]} \BibitemShut
  {NoStop}%
\bibitem [{\citenamefont {Abelev}\ \emph {et~al.}(2013)\citenamefont {Abelev}
  \emph {et~al.}}]{Abelev:2012ba}%
  \BibitemOpen
  \bibfield  {author} {\bibinfo {author} {\bibfnamefont {B.}~\bibnamefont
  {Abelev}} \emph {et~al.} (\bibinfo {collaboration} {ALICE}),\ }\href
  {\doibase 10.1016/j.physletb.2012.11.059} {\bibfield  {journal} {\bibinfo
  {journal} {Phys. Lett. B}\ }\textbf {\bibinfo {volume} {718}},\ \bibinfo
  {pages} {1273} (\bibinfo {year} {2013})},\ \Eprint
  {http://arxiv.org/abs/1209.3715} {arXiv:1209.3715 [nucl-ex]} \BibitemShut
  {NoStop}%
\bibitem [{\citenamefont {Adam}\ \emph {et~al.}(2015)\citenamefont {Adam} \emph
  {et~al.}}]{Adam:2015gsa}%
  \BibitemOpen
  \bibfield  {author} {\bibinfo {author} {\bibfnamefont {J.}~\bibnamefont
  {Adam}} \emph {et~al.} (\bibinfo {collaboration} {ALICE}),\ }\href {\doibase
  10.1007/JHEP09(2015)095} {\bibfield  {journal} {\bibinfo  {journal} {JHEP}\
  }\textbf {\bibinfo {volume} {09}},\ \bibinfo {pages} {095} (\bibinfo {year}
  {2015})},\ \Eprint {http://arxiv.org/abs/1503.09177} {arXiv:1503.09177
  [nucl-ex]} \BibitemShut {NoStop}%
\bibitem [{\citenamefont {Adamczyk}\ \emph {et~al.}(2017)\citenamefont
  {Adamczyk} \emph {et~al.}}]{Adamczyk:2017vfu}%
  \BibitemOpen
  \bibfield  {author} {\bibinfo {author} {\bibfnamefont {L.}~\bibnamefont
  {Adamczyk}} \emph {et~al.} (\bibinfo {collaboration} {STAR}),\ }\href
  {\doibase 10.1103/PhysRevC.96.054904} {\bibfield  {journal} {\bibinfo
  {journal} {Phys. Rev. C}\ }\textbf {\bibinfo {volume} {96}},\ \bibinfo
  {pages} {054904} (\bibinfo {year} {2017})},\ \Eprint
  {http://arxiv.org/abs/1702.07705} {arXiv:1702.07705 [nucl-ex]} \BibitemShut
  {NoStop}%
\bibitem [{\citenamefont {Acharya}\ \emph {et~al.}(2020)\citenamefont {Acharya}
  \emph {et~al.}}]{Acharya:2020sbc}%
  \BibitemOpen
  \bibfield  {author} {\bibinfo {author} {\bibfnamefont {S.}~\bibnamefont
  {Acharya}} \emph {et~al.} (\bibinfo {collaboration} {ALICE}),\ }\href
  {\doibase 10.1007/JHEP06(2020)035} {\bibfield  {journal} {\bibinfo  {journal}
  {JHEP}\ }\textbf {\bibinfo {volume} {06}},\ \bibinfo {pages} {035} (\bibinfo
  {year} {2020})},\ \Eprint {http://arxiv.org/abs/2002.10897} {arXiv:2002.10897
  [nucl-ex]} \BibitemShut {NoStop}%
\bibitem [{\citenamefont {Acharya}\ \emph
  {et~al.}(2021{\natexlab{a}})\citenamefont {Acharya} \emph
  {et~al.}}]{ALICE:2021jnv}%
  \BibitemOpen
  \bibfield  {author} {\bibinfo {author} {\bibfnamefont {S.}~\bibnamefont
  {Acharya}} \emph {et~al.} (\bibinfo {collaboration} {ALICE}),\ }\href
  {\doibase 10.1016/j.physletb.2021.136481} {\bibfield  {journal} {\bibinfo
  {journal} {Phys. Lett. B}\ }\textbf {\bibinfo {volume} {820}},\ \bibinfo
  {pages} {136481} (\bibinfo {year} {2021}{\natexlab{a}})},\ \Eprint
  {http://arxiv.org/abs/2101.02581} {arXiv:2101.02581 [nucl-ex]} \BibitemShut
  {NoStop}%
\bibitem [{\citenamefont {Acharya}\ \emph
  {et~al.}(2021{\natexlab{b}})\citenamefont {Acharya} \emph
  {et~al.}}]{ALICE:2021tyx}%
  \BibitemOpen
  \bibfield  {author} {\bibinfo {author} {\bibfnamefont {S.}~\bibnamefont
  {Acharya}} \emph {et~al.} (\bibinfo {collaboration} {ALICE}),\ }\href
  {\doibase 10.1016/j.physletb.2021.136280} {\bibfield  {journal} {\bibinfo
  {journal} {Phys. Lett. B}\ }\textbf {\bibinfo {volume} {817}},\ \bibinfo
  {pages} {136280} (\bibinfo {year} {2021}{\natexlab{b}})},\ \Eprint
  {http://arxiv.org/abs/2101.04623} {arXiv:2101.04623 [nucl-ex]} \BibitemShut
  {NoStop}%
\bibitem [{\citenamefont {Acharya}\ \emph
  {et~al.}(2021{\natexlab{c}})\citenamefont {Acharya} \emph
  {et~al.}}]{ALICE:2021gpt}%
  \BibitemOpen
  \bibfield  {author} {\bibinfo {author} {\bibfnamefont {S.}~\bibnamefont
  {Acharya}} \emph {et~al.} (\bibinfo {collaboration} {ALICE}),\ }\href@noop {}
  {\  (\bibinfo {year} {2021}{\natexlab{c}})},\ \Eprint
  {http://arxiv.org/abs/2101.04577} {arXiv:2101.04577 [nucl-ex]} \BibitemShut
  {NoStop}%
\bibitem [{\citenamefont {Aschenauer}\ \emph {et~al.}(2019)\citenamefont
  {Aschenauer}, \citenamefont {Baker}, \citenamefont {Chang}, \citenamefont
  {Lee}, \citenamefont {Tu},\ and\ \citenamefont {Zheng}}]{Beagle}%
  \BibitemOpen
  \bibfield  {author} {\bibinfo {author} {\bibfnamefont {E.}~\bibnamefont
  {Aschenauer}}, \bibinfo {author} {\bibfnamefont {M.}~\bibnamefont {Baker}},
  \bibinfo {author} {\bibfnamefont {W.}~\bibnamefont {Chang}}, \bibinfo
  {author} {\bibfnamefont {J.}~\bibnamefont {Lee}}, \bibinfo {author}
  {\bibfnamefont {Z.}~\bibnamefont {Tu}}, \ and\ \bibinfo {author}
  {\bibfnamefont {L.}~\bibnamefont {Zheng}},\ }\href@noop {} {\enquote
  {\bibinfo {title} {{BeAGLE: A Tool to Refine Detector Requirements for eA
  Collisions EIC R\&D Project eRD17: Progress Report (January-June 2019) and
  Proposal}},}\ }\bibinfo {howpublished} {\url
  {https://wiki.bnl.gov/eic/index.php/BeAGLE}} (\bibinfo {year}
  {2019})\BibitemShut {NoStop}%
\bibitem [{\citenamefont {Sj\"ostrand}\ \emph {et~al.}(2006)\citenamefont
  {Sj\"ostrand}, \citenamefont {Mrenna},\ and\ \citenamefont
  {Skands}}]{Sjostrand:2006za}%
  \BibitemOpen
  \bibfield  {author} {\bibinfo {author} {\bibfnamefont {T.}~\bibnamefont
  {Sj\"ostrand}}, \bibinfo {author} {\bibfnamefont {S.}~\bibnamefont {Mrenna}},
  \ and\ \bibinfo {author} {\bibfnamefont {P.}~\bibnamefont {Skands}},\ }\href
  {\doibase 10.1088/1126-6708/2006/05/026} {\bibfield  {journal} {\bibinfo
  {journal} {JHEP}\ }\textbf {\bibinfo {volume} {05}},\ \bibinfo {pages} {026}
  (\bibinfo {year} {2006})},\ \Eprint {http://arxiv.org/abs/hep-ph/0603175}
  {arXiv:hep-ph/0603175 [hep-ph]} \BibitemShut {NoStop}%
\bibitem [{\citenamefont {Roesler}\ \emph {et~al.}(2000)\citenamefont
  {Roesler}, \citenamefont {Engel},\ and\ \citenamefont
  {Ranft}}]{Roesler:2000he}%
  \BibitemOpen
  \bibfield  {author} {\bibinfo {author} {\bibfnamefont {S.}~\bibnamefont
  {Roesler}}, \bibinfo {author} {\bibfnamefont {R.}~\bibnamefont {Engel}}, \
  and\ \bibinfo {author} {\bibfnamefont {J.}~\bibnamefont {Ranft}},\ }in\ \href
  {\doibase 10.1007/978-3-642-18211-2\_166} {\emph {\bibinfo {booktitle}
  {{Advanced Monte Carlo for radiation physics, particle transport simulation
  and applications. Proceedings, Conference, MC2000, Lisbon, Portugal, October
  23-26, 2000}}}}\ (\bibinfo {year} {2000})\ pp.\ \bibinfo {pages}
  {1033--1038},\ \Eprint {http://arxiv.org/abs/hep-ph/0012252}
  {arXiv:hep-ph/0012252} \BibitemShut {NoStop}%
\bibitem [{\citenamefont {Böhlen}\ \emph {et~al.}(2014)\citenamefont
  {Böhlen}, \citenamefont {Cerutti}, \citenamefont {Chin}, \citenamefont
  {Fassò}, \citenamefont {Ferrari}, \citenamefont {Ortega}, \citenamefont
  {Mairani}, \citenamefont {Sala}, \citenamefont {Smirnov},\ and\ \citenamefont
  {Vlachoudis}}]{Bohlen:2014buj}%
  \BibitemOpen
  \bibfield  {author} {\bibinfo {author} {\bibfnamefont {T.}~\bibnamefont
  {Böhlen}}, \bibinfo {author} {\bibfnamefont {F.}~\bibnamefont {Cerutti}},
  \bibinfo {author} {\bibfnamefont {M.}~\bibnamefont {Chin}}, \bibinfo {author}
  {\bibfnamefont {A.}~\bibnamefont {Fassò}}, \bibinfo {author} {\bibfnamefont
  {A.}~\bibnamefont {Ferrari}}, \bibinfo {author} {\bibfnamefont
  {P.}~\bibnamefont {Ortega}}, \bibinfo {author} {\bibfnamefont
  {A.}~\bibnamefont {Mairani}}, \bibinfo {author} {\bibfnamefont
  {P.}~\bibnamefont {Sala}}, \bibinfo {author} {\bibfnamefont {G.}~\bibnamefont
  {Smirnov}}, \ and\ \bibinfo {author} {\bibfnamefont {V.}~\bibnamefont
  {Vlachoudis}},\ }\href {\doibase 10.1016/j.nds.2014.07.049} {\bibfield
  {journal} {\bibinfo  {journal} {Nucl. Data Sheets}\ }\textbf {\bibinfo
  {volume} {120}},\ \bibinfo {pages} {211} (\bibinfo {year}
  {2014})}\BibitemShut {NoStop}%
\bibitem [{\citenamefont {Ferrari}\ \emph {et~al.}(2005)\citenamefont
  {Ferrari}, \citenamefont {Sala}, \citenamefont {Fasso},\ and\ \citenamefont
  {Ranft}}]{Ferrari:2005zk}%
  \BibitemOpen
  \bibfield  {author} {\bibinfo {author} {\bibfnamefont {A.}~\bibnamefont
  {Ferrari}}, \bibinfo {author} {\bibfnamefont {P.~R.}\ \bibnamefont {Sala}},
  \bibinfo {author} {\bibfnamefont {A.}~\bibnamefont {Fasso}}, \ and\ \bibinfo
  {author} {\bibfnamefont {J.}~\bibnamefont {Ranft}},\ }\href {\doibase
  10.2172/877507} {\  (\bibinfo {year} {2005}),\ 10.2172/877507}\BibitemShut
  {NoStop}%
\bibitem [{\citenamefont {Tu}\ \emph {et~al.}(2020)\citenamefont {Tu},
  \citenamefont {Jentsch}, \citenamefont {Baker}, \citenamefont {Zheng},
  \citenamefont {Lee}, \citenamefont {Venugopalan}, \citenamefont {Hen},
  \citenamefont {Higinbotham}, \citenamefont {Aschenauer},\ and\ \citenamefont
  {Ullrich}}]{Tu:2020ymk}%
  \BibitemOpen
  \bibfield  {author} {\bibinfo {author} {\bibfnamefont {Z.}~\bibnamefont
  {Tu}}, \bibinfo {author} {\bibfnamefont {A.}~\bibnamefont {Jentsch}},
  \bibinfo {author} {\bibfnamefont {M.}~\bibnamefont {Baker}}, \bibinfo
  {author} {\bibfnamefont {L.}~\bibnamefont {Zheng}}, \bibinfo {author}
  {\bibfnamefont {J.-H.}\ \bibnamefont {Lee}}, \bibinfo {author} {\bibfnamefont
  {R.}~\bibnamefont {Venugopalan}}, \bibinfo {author} {\bibfnamefont
  {O.}~\bibnamefont {Hen}}, \bibinfo {author} {\bibfnamefont {D.}~\bibnamefont
  {Higinbotham}}, \bibinfo {author} {\bibfnamefont {E.-C.}\ \bibnamefont
  {Aschenauer}}, \ and\ \bibinfo {author} {\bibfnamefont {T.}~\bibnamefont
  {Ullrich}},\ }\href {\doibase 10.1016/j.physletb.2020.135877} {\bibfield
  {journal} {\bibinfo  {journal} {Phys. Lett. B}\ }\textbf {\bibinfo {volume}
  {811}},\ \bibinfo {pages} {135877} (\bibinfo {year} {2020})},\ \Eprint
  {http://arxiv.org/abs/2005.14706} {arXiv:2005.14706 [nucl-ex]} \BibitemShut
  {NoStop}%
\bibitem [{\citenamefont {Bertini}(1963)}]{PhysRev.131.1801}%
  \BibitemOpen
  \bibfield  {author} {\bibinfo {author} {\bibfnamefont {H.~W.}\ \bibnamefont
  {Bertini}},\ }\href {\doibase 10.1103/PhysRev.131.1801} {\bibfield  {journal}
  {\bibinfo  {journal} {Phys. Rev.}\ }\textbf {\bibinfo {volume} {131}},\
  \bibinfo {pages} {1801} (\bibinfo {year} {1963})}\BibitemShut {NoStop}%
\bibitem [{\citenamefont {Ferrari}\ \emph {et~al.}(1996)\citenamefont
  {Ferrari}, \citenamefont {Sala}, \citenamefont {Ranft},\ and\ \citenamefont
  {Roesler}}]{Ferrari:1995cq}%
  \BibitemOpen
  \bibfield  {author} {\bibinfo {author} {\bibfnamefont {A.}~\bibnamefont
  {Ferrari}}, \bibinfo {author} {\bibfnamefont {P.~R.}\ \bibnamefont {Sala}},
  \bibinfo {author} {\bibfnamefont {J.}~\bibnamefont {Ranft}}, \ and\ \bibinfo
  {author} {\bibfnamefont {S.}~\bibnamefont {Roesler}},\ }\href {\doibase
  10.1007/s002880050119} {\bibfield  {journal} {\bibinfo  {journal} {Z. Phys.
  C}\ }\textbf {\bibinfo {volume} {70}},\ \bibinfo {pages} {413} (\bibinfo
  {year} {1996})},\ \Eprint {http://arxiv.org/abs/nucl-th/9509039}
  {arXiv:nucl-th/9509039} \BibitemShut {NoStop}%
\bibitem [{\citenamefont {Zheng}\ \emph {et~al.}(2014)\citenamefont {Zheng},
  \citenamefont {Aschenauer},\ and\ \citenamefont {Lee}}]{Zheng:2014cha}%
  \BibitemOpen
  \bibfield  {author} {\bibinfo {author} {\bibfnamefont {L.}~\bibnamefont
  {Zheng}}, \bibinfo {author} {\bibfnamefont {E.}~\bibnamefont {Aschenauer}}, \
  and\ \bibinfo {author} {\bibfnamefont {J.}~\bibnamefont {Lee}},\ }\href
  {\doibase 10.1140/epja/i2014-14189-3} {\bibfield  {journal} {\bibinfo
  {journal} {Eur. Phys. J. A}\ }\textbf {\bibinfo {volume} {50}},\ \bibinfo
  {pages} {189} (\bibinfo {year} {2014})},\ \Eprint
  {http://arxiv.org/abs/1407.8055} {arXiv:1407.8055 [hep-ex]} \BibitemShut
  {NoStop}%
\bibitem [{\citenamefont {Adams}\ \emph {et~al.}(1995)\citenamefont {Adams}
  \emph {et~al.}}]{E665:1995utr}%
  \BibitemOpen
  \bibfield  {author} {\bibinfo {author} {\bibfnamefont {M.~R.}\ \bibnamefont
  {Adams}} \emph {et~al.} (\bibinfo {collaboration} {E665}),\ }\href {\doibase
  10.1103/PhysRevLett.74.5198} {\bibfield  {journal} {\bibinfo  {journal}
  {Phys. Rev. Lett.}\ }\textbf {\bibinfo {volume} {74}},\ \bibinfo {pages}
  {5198} (\bibinfo {year} {1995})},\ \bibinfo {note} {[Erratum: Phys.Rev.Lett.
  80, 2020--2021 (1998)]}\BibitemShut {NoStop}%
\bibitem [{\citenamefont {Frankfurt}\ and\ \citenamefont
  {Strikman}(1988)}]{Frankfurt:1988nt}%
  \BibitemOpen
  \bibfield  {author} {\bibinfo {author} {\bibfnamefont {L.~L.}\ \bibnamefont
  {Frankfurt}}\ and\ \bibinfo {author} {\bibfnamefont {M.~I.}\ \bibnamefont
  {Strikman}},\ }\href {\doibase 10.1016/0370-1573(88)90179-2} {\bibfield
  {journal} {\bibinfo  {journal} {Phys. Rept.}\ }\textbf {\bibinfo {volume}
  {160}},\ \bibinfo {pages} {235} (\bibinfo {year} {1988})}\BibitemShut
  {NoStop}%
\bibitem [{\citenamefont {Eskola}\ \emph {et~al.}(2009)\citenamefont {Eskola},
  \citenamefont {Paukkunen},\ and\ \citenamefont {Salgado}}]{Eskola:2009uj}%
  \BibitemOpen
  \bibfield  {author} {\bibinfo {author} {\bibfnamefont {K.}~\bibnamefont
  {Eskola}}, \bibinfo {author} {\bibfnamefont {H.}~\bibnamefont {Paukkunen}}, \
  and\ \bibinfo {author} {\bibfnamefont {C.}~\bibnamefont {Salgado}},\ }\href
  {\doibase 10.1088/1126-6708/2009/04/065} {\bibfield  {journal} {\bibinfo
  {journal} {JHEP}\ }\textbf {\bibinfo {volume} {04}},\ \bibinfo {pages} {065}
  (\bibinfo {year} {2009})},\ \Eprint {http://arxiv.org/abs/0902.4154}
  {arXiv:0902.4154 [hep-ph]} \BibitemShut {NoStop}%
\bibitem [{\citenamefont {Kiselev}\ and\ \citenamefont
  {Jentsch}(2020)}]{ref:EICROOT}%
  \BibitemOpen
  \bibfield  {author} {\bibinfo {author} {\bibfnamefont {A.}~\bibnamefont
  {Kiselev}}\ and\ \bibinfo {author} {\bibfnamefont {A.}~\bibnamefont
  {Jentsch}},\ }\href@noop {} {\enquote {\bibinfo {title} {Eic root: A
  light-weight geant detector simulation software suite based on the fairroot
  framework.}}\ }\bibinfo {howpublished} {\url{https://github.com/eic/EicRoot}}
  (\bibinfo {year} {2020})\BibitemShut {NoStop}%
\bibitem [{\citenamefont {Brun}\ \emph {et~al.}(1987)\citenamefont {Brun},
  \citenamefont {Bruyant}, \citenamefont {Maire}, \citenamefont {McPherson},\
  and\ \citenamefont {Zanarini}}]{Brun:1987ma}%
  \BibitemOpen
  \bibfield  {author} {\bibinfo {author} {\bibfnamefont {R.}~\bibnamefont
  {Brun}}, \bibinfo {author} {\bibfnamefont {F.}~\bibnamefont {Bruyant}},
  \bibinfo {author} {\bibfnamefont {M.}~\bibnamefont {Maire}}, \bibinfo
  {author} {\bibfnamefont {A.~C.}\ \bibnamefont {McPherson}}, \ and\ \bibinfo
  {author} {\bibfnamefont {P.}~\bibnamefont {Zanarini}},\ }\href@noop {}
  {\enquote {\bibinfo {title} {{GEANT3}},}\ }\bibinfo {howpublished}
  {\url{https://cds.cern.ch/record/1119728}} (\bibinfo {year}
  {1987})\BibitemShut {NoStop}%
\bibitem [{\citenamefont {Agostinelli}\ \emph {et~al.}(2003)\citenamefont
  {Agostinelli} \emph {et~al.}}]{GEANT4}%
  \BibitemOpen
  \bibfield  {author} {\bibinfo {author} {\bibfnamefont {S.}~\bibnamefont
  {Agostinelli}} \emph {et~al.} (\bibinfo {collaboration} {GEANT4}),\ }\href
  {\doibase 10.1016/S0168-9002(03)01368-8} {\bibfield  {journal} {\bibinfo
  {journal} {Nucl. Instrum. Meth. A}\ }\textbf {\bibinfo {volume} {506}},\
  \bibinfo {pages} {250} (\bibinfo {year} {2003})}\BibitemShut {NoStop}%
\end{thebibliography}%
\end{document}